\documentclass{amsart}
\usepackage{graphicx}
\usepackage{amsmath}
\usepackage{lipsum}
\usepackage[varg]{txfonts}
\usepackage{here}
\usepackage{setspace}
\begin{document}
\begin{center}
\begin{spacing}{2}
\begin{Huge}
Experimental Study on Battery-less Sensor Network Activated by Multi-point Wireless Energy Transmission
\end{Huge}
\end{spacing}
\end{center}

\begin{large}
\begin{center}
Daiki Maehara$^{1}$, {Gia Khanh} Tran$^{1}$, Kei Sakaguchi$^{1}$, Kiyomichi Araki$^{1}$  \\ 
\begin{spacing}{2}
\end{spacing}
$^{1}$Tokyo Institute of Technology, Tokyo, Japan \\
Email: {maehara@mobile.ee.titech.ac.jp} \\
\end{center}
\end{large}
\begin{spacing}{2}
\end{spacing}
\begin{flushright}
DRAFT: December 2015
\end{flushright}

\begin{abstract}
This paper empirically validates battery-less sensor activation via wireless energy transmission to release sensors from wires and batteries. To seamlessly extend the coverage and activate sensor nodes distributed in any indoor environment, we proposed multi-point wireless energy transmission with carrier shift diversity. In this scheme, multiple transmitters are employed to compensate path-loss attenuation and orthogonal frequencies are allocated to the multiple transmitters to avoid the destructive interference that occurs when the same frequency is used by all transmitters. In our previous works, the effectiveness of the proposed scheme was validated theoretically and also empirically by using just a spectrum analyzer to measure the received power. In this paper, we develop low-energy battery-less sensor nodes whose consumed power and required received power for activation are respectively 142~$\mu$W and 400~$\mu$W. In addition, we conduct indoor experiments in which the received power and activation of battery-less sensor node are simultaneously observed by using the developed battery-less sensor node and a spectrum analyzer. The results show that the coverage of single-point and multi-point wireless energy transmission without carrier shift diversity are, respectively, 84.4\% and 83.7\%, while the coverage of the proposed scheme is 100\%. It can be concluded that the effectiveness of the proposed scheme can be verified by our experiments using real battery-less sensor nodes.
\end{abstract}

\section{Introduction}
\label{sec:Intro}
Wireless Sensor Networks (WSNs) employ numerous sensors deployed in target environments to gather sensing data via wireless links. In the conventional WSNs, installation placement of sensor nodes operated by electrical plug are restricted by limited placements of the plugs. Employing electrical batteries instead can release WSNs from the wired cables however increase the cost for battery replacement.  Therefore, the demand of battery-less sensor nodes has been increasing in WSNs. One of the battery-less solutions is energy harvesting from light source\cite{solar}, RF (Radio Frequency) \cite{RF}, vibration \cite{vib} and so on. However, the ambient energy source is easily influenced by the surrounding environment. For example, light source can only be employed in the daytime and vibration source needs to be triggered by the external force e.g. pushing by human finger.

To overcome the power supply problem, a system called wireless grid to supply wireless energy to sensor nodes has been proposed \cite{Paper0}. Wireless energy transmission schemes are categorized into three types, i.e. radio wave emission, resonant coupling, and inductive coupling \cite{book1}. In the radio wave emission method,  power emitted from the wireless energy transmitter attenuates in proportion to the distance. However, the attenuation can be compensated by enhancing antenna gain. The most important advantage of this scheme is that it can supply energy to unrestricted numbers of sensors without any strict knowledge of sensor locations. Since our study focuses on supplying wireless energy to ubiquitous sensor nodes distributed in indoor environments, the radio wave emission approach is employed. 

Radio wave emission technology has long been employed in RFID (Radio Frequency IDentification) systems to manage products in the factories. However, the wireless energy transmission systems were not designed to achieve the ubiquitous coverage demanded of the wireless grid. In single transmitter (Tx) systems, the coverage of energy supply field is restricted by the maximum transmit power or EIRP (Equivalent Isotropic Radiated Power) defined by radio regulation. To extend the coverage, multiple Txs are introduced to the systems. However, interference can occur between multiple Txs. Once multiple Txs simultaneously transmit radio wave, destructive interference occurs and creates deadspots where sensor nodes cannot be charged and activated. To deal with this problem, we proposed to introduce Carrier Shift Diversity (CSD) to the multi-point system \cite{Paper1}\cite{ISSSE}. Since the CSD creates artificial fading and the instantaneous received power fluctuates against the required power for activating sensor node, capacitor playing the role of rechargeable battery is introduced to average the received power over the period of artificial fading. 

In our previous works \cite{Paper1}\cite{ISSSE}, we conducted indoor experiments performed in a three dimensional space of an empty room by measuring only the propagation characteristics using a receiving antenna of IC tag and spectrum analyzer. The experimental results showed that the multi-point wireless energy transmission with CSD can mitigate power attenuation due to the path-loss as well as the effect of standing-wave created by multipath and interference between multiple wave sources. However, in these papers, the coverage of sensor activation was not confirmed with real battery-less sensor nodes. In the scenario of real battery-less sensor node, RF/DC conversion efficiency is non-linear against received power and the consumed power varies according to the activation status of the sensor node. Therefore, \cite{Paper1}\cite{ISSSE} could not prove the effectiveness of the proposed scheme in terms of activating real battery-less sensor nodes. In this paper, we develop battery-less sensor node composed of off-the-shelf devices and conduct indoor experiments by using the developed sensor node. In the experiments, we measure both received power and sensor activation using the developed battery-less sensor node whose consumed power and required power to activate sensor node are respectively $140\mu$W and $400\mu$W. The experimental results show that the coverage of sensor activation of the single-point scheme, the simple multi-point scheme without CSD, and the proposed multi-point scheme are 84.4\%, 83.7\%, and 100\% respectively. In this paper, the details of the developed battery-less sensor node, measurement scheme, and measurement results are given in addition to the contents in our previous report \cite{PIMRC}.

For the rest of this paper, Sec.~\ref{sec:MPC} revisits the concept of multi-point wireless energy transmission with CSD. Section \ref{sec:BLS} describes the design concept of a low-energy battery-less sensor node and presents the developed prototype hardware with measured power consumption. Section~~\ref{sec:Ex} shows the indoor experiment of the proposed wireless energy transmission scheme with the prototype hardware. Finally, Sec.~\ref{sec:concl} concludes this paper. In this paper, single-point, simple multi-point and proposed multi-point are abbreviated as SP, MP and MPCSD respectively.

\begin{figure}[!t]
\centering 
\includegraphics[width=5in]{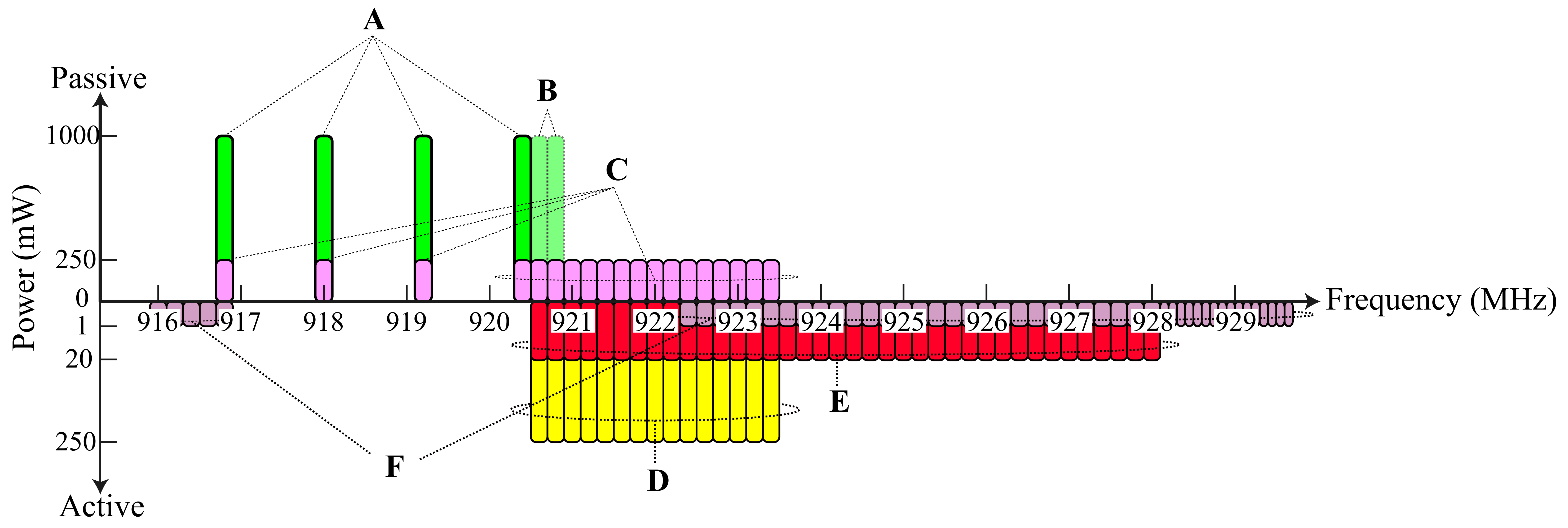}
%\includepdf[width=5in]{figure/chapter2/920_mask.pdf}
\caption{Radio regulation in 920~MHz band.}
\label{fig:920}
\end{figure}

\begin{figure}[!t]
\centering
\includegraphics[width=5in]{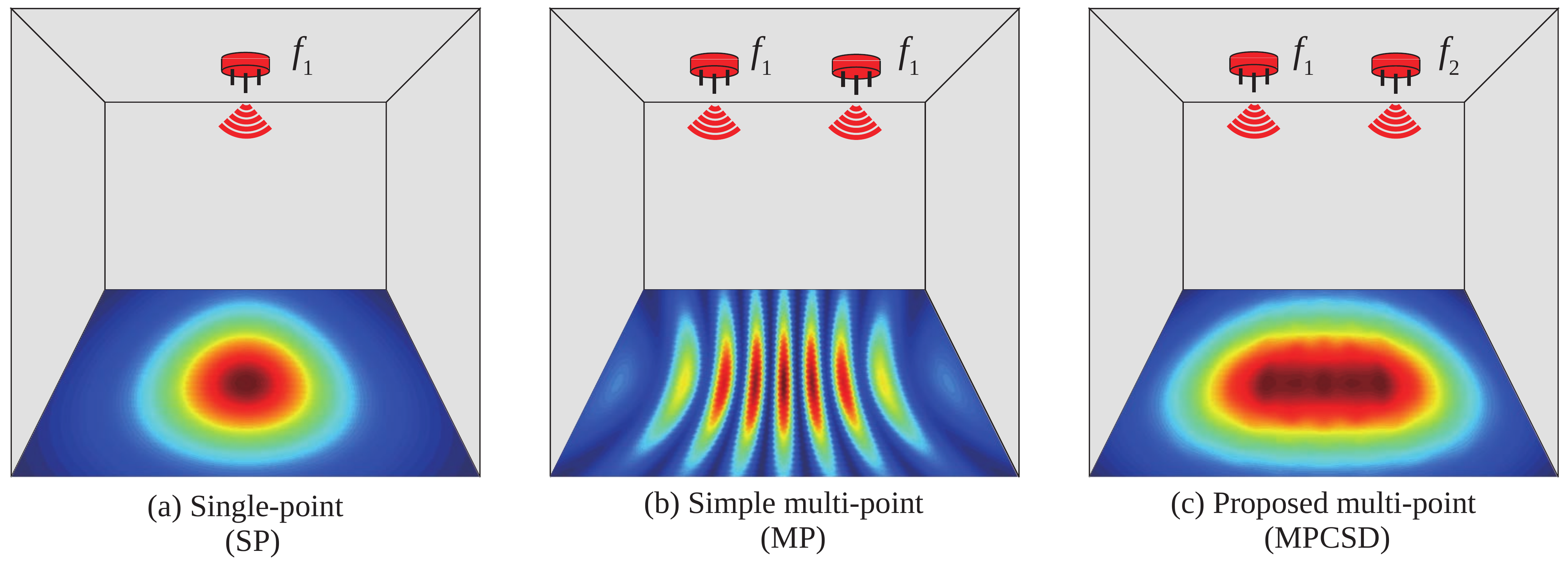}
\caption{Concept of multi-point wireless energy transmission.}
\label{fig:MPC}
\end{figure}

\section{Multi-point wireless energy transmission with carrier shift diversity}
\label{sec:MPC}
This section introduces 920~MHz band which is used for wireless energy transmission and wireless sensor networks in Japan, revisits the concept of MPCSD compared with SP or MP in terms of the coverage of energy supply field, and provides theoretical discussion of the coverage by considering the non-linearity of RF/DC conversion efficiency and the time-varying received power.

\subsection{920MHz band}
Radio wave emission method for energy transmission is regulated at 920~MHz band in Japan.  Figure~\ref{fig:920} shows the spectrum mask at 920~MHz band \cite{Reg1}. According to Japan's radio regulation, 920~MHz band is categorized into 6 systems as shown in Fig.~\ref{fig:920}. $\bf{A}$ and $\bf{B}$ are used for passive tag systems, where the maximum transmit power and the maximum EIRP are limited by 1~W (30~dBm) and 4~W (36~dBm) respectively. In addition, at $\bf{A}$, LBT (Listen Before Talk), which is a scheme of carrier sensing, is not required. $\bf{C}$ can be used for specified low power radio station of passive tag systems. At $\bf{C}$, the maximum transmit power is limited by 250~mW (24~dBm). In these passive tag systems, channels, where center frequencies are 915.8~MHz to 921.4~MHz, can be used for the backscatter signals from the passive tags.   $\bf{D}$ can be used for convenience radio station of active tag systems, where the maximum transmit power is limited by 250~mW (24~dBm). $\bf{E}$ and $\bf{F}$ can be used for  specified low power radio station, where the maximum transmit powers are limited by 20~mW (13~dBm) and 1~mW (0~dBm) respectively. In the proposed wireless grid,  four channels at $\bf{A}$ are employed for wireless energy transmission to perform continuous energy supply, while the channels at $\bf{F}$ are used for data communication.

\subsection{Multi-point Wireless Energy Transmission}
Figure \ref{fig:MPC} shows the concept of three energy transmission schemes. In SP, the coverage of energy supply is restricted by the maximum transmit power as shown in Fig. \ref{fig:MPC}(a). In free-space condition, the received power decreases in proportion to the square of the distance from Tx. 

\begin{figure}[!t]
\centering
\includegraphics[width=3in]{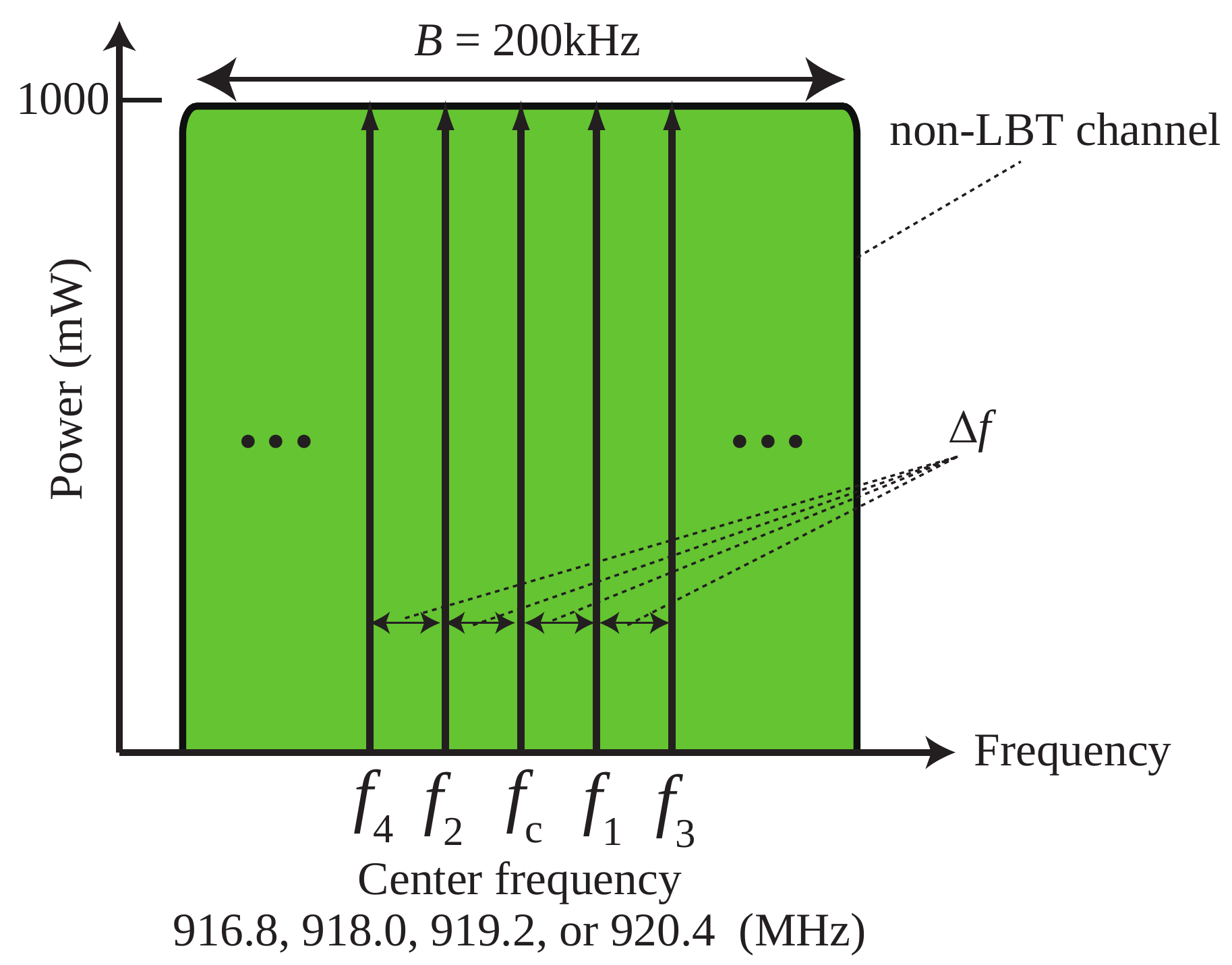}
\caption{Concept of carrier shift diversity.}
\label{fig:CSC}
\end{figure}

To enhance the area of energy supply field, multiple Txs can be added to the systems. However, deadspot is created due to the destructive interference that occurs when the same frequency is used by all transmitters as shown in Fig. \ref{fig:MPC}(b). To combat the deadspot problem, we proposed to apply CSD to the multi-point scheme. By this scheme, the destructive interference can be canceled out in time averaging so that the coverage can be seamlessly extended as shown in Fig. \ref{fig:MPC}(c).  To perform the CSD, when there are $N$ Txs supplying energy to sensor nodes, 200~kHz of the available frequency band is divided into $N$ orthogonal subcarriers as shown in Fig. \ref{fig:CSC} at an interval of $\Delta f$ by which the cycle of artificial fading can be configured as $T_\mathrm{f}=1/\Delta f$. In addition, sensor node has a capacitor which plays a role of averaging its received power fluctuated by the artificial fading. It is noted that the frequency bandwidth in the case of CSD is as same as that in the cases of SP and MP because CSD merely divides the original dedicated channel into the subcarriers.

\subsection{Estimation of energy transmission coverage}
\label{sec:Est}
\begin{figure}[!t]
\centering
\includegraphics[width=3in]{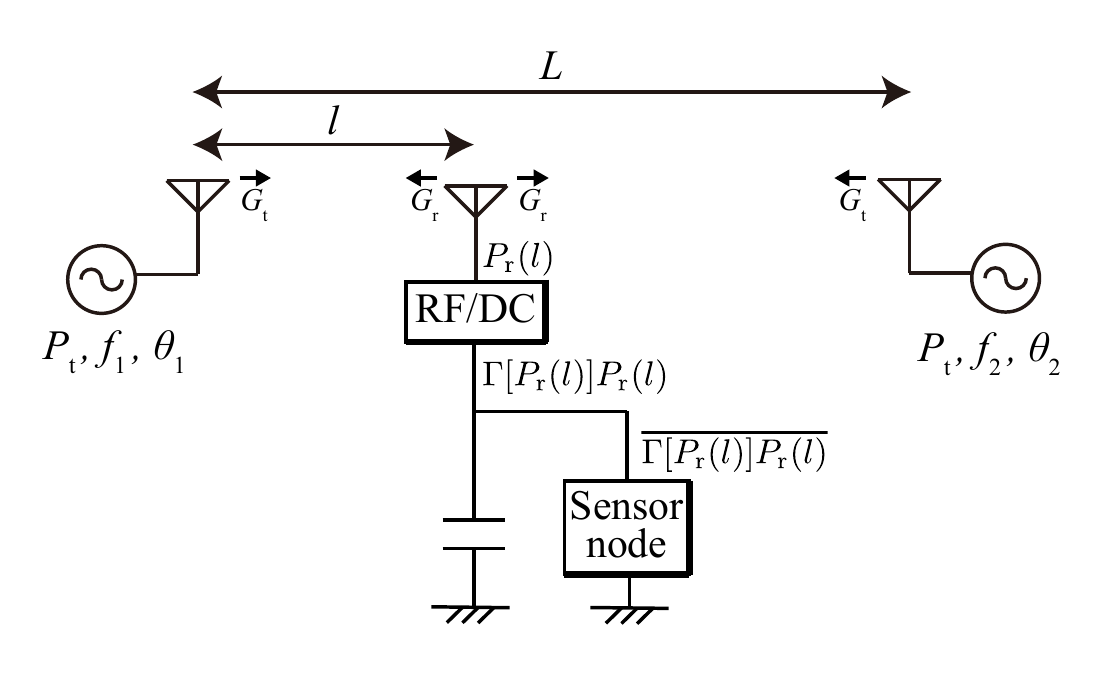}
\caption{Two transmitter model.}
\label{fig:two}
\end{figure}

In order to understand the effectiveness of the proposed multi-point scheme, this subsection provides theoretical discussion on the energy transmission coverage. As shown in Fig. \ref{fig:two}, a model of two transmitters separated with a distance $L$ is assumed, which is almost in the same condition with the experiments presented in Sect.~\ref{sec:Ex}. In the theoretical discussions, $P_\mathrm{t}$, $G_\mathrm{t}$ and $G_\mathrm{r}$ show the transmit power, the transmit antenna gain of energy transmitters and the receive antenna gain of the sensor nodes respectively. For the sake of brevity, this paper defines $P_\mathrm{t}^\mathrm{e}=P_\mathrm{t}G_\mathrm{t}G_\mathrm{r}$ as the equivalent transmit power used in the consequent formulas since the model only considers 1-D space. In addition, $f_1$ and $f_2$ show the center frequencies of Tx$\sharp$1 and Tx$\sharp$2 respectively. $\theta_1$ and $\theta_2$ show the initial phase of Tx$\sharp$1 and Tx$\sharp$2 respectively, and $\Delta \theta = \theta_1-\theta_2$ denotes phase difference between two transmitters. In the case of SP, single transmitter of either Tx$\sharp$1 or Tx$\sharp$2 is employed. In the case of MP, both Tx$\sharp$1 and Tx$\sharp$2, whose center frequencies are the same ($f_2=f_1$), are employed, while the carrier offset $\Delta f$ is employed ($f_2 = f_1 + \Delta f$) in the case of MPCSD.  As described in the previous section, the MPCSD creates artificial fading between the two transmitters with the cycle of $T_\mathrm{f} = 1/ \Delta f$.

The RF received power $P_\mathrm{r}(l)$ at location $l$ is converted into DC power as $\Gamma[P_\mathrm{r}(l)]P_\mathrm{r}(l)$ where $\Gamma[P_\mathrm{r}(l)]$ is the RF/DC conversion efficiency at the RF received power $P_\mathrm{r}(l)$. On the other hand, $P_\mathrm{csp}$ is the average consumed power of sensor node over the duty cycle $T_\mathrm{d}$. It is noted that the consumed power is not constant against time because the sensor node employs an intermittent operation whose details will be presented in Sect.~\ref{sec:BLS}. Since the MPCSD also creates fluctuation of received power over the cycle of $T_\mathrm{f}$, the received power is averaged to $\overline{\Gamma[P_\mathrm{r}(l)]P_\mathrm{r}(l)}$ by a capacitor. Because the average received power should be higher than the average consumed power of the sensor node $P_\mathrm{csp}$ to permanently activate the sensor node, the activation condition $\tilde{A}$ can be defined as
\begin{eqnarray}
\tilde{A}(l)= \begin{cases}
1 & \mathrm{if} \ \ \overline{\Gamma[P_\mathrm{r}(l)]P_\mathrm{r}(l)}\geq P_\mathrm{csp} \\
0 & \mathrm{if} \ \ \overline{\Gamma[P_\mathrm{r}(l)]P_\mathrm{r}(l)} < P_\mathrm{csp}.
\end{cases}
\label{equ:ff}
\end{eqnarray}
For further simplification of Eq.~(\ref{equ:ff}), we will introduce the low power operation of RF/DC conversion circuit \cite{Chris}, where $\partial^2 \left( \Gamma[P_\mathrm{r}(l)] P_\mathrm{r}(l) \right)/\partial {P_\mathrm{r}(l)}^2 \ge 0$ can be assumed since threshold voltage of the diodes is dominant for the efficiency in low power operation. If this assumption is satisfied, the following inequality holds in accordance with the Jensen's inequality, 
\begin{eqnarray}
\label{equ:approx}
\Gamma[\overline{P_\mathrm{r}(l)}]\overline{P_\mathrm{r}(l)} \leq \overline{\Gamma[P_\mathrm{r}(l)]P_\mathrm{r}(l)}.
\end{eqnarray}
Therefore, the activation condition in Eq.~(\ref{equ:ff}) can be rewritten as
\begin{eqnarray}
A(l)= \begin{cases}
1 & \mathrm{if} \ \ \Gamma[\overline{P_\mathrm{r}(l)}]\overline{P_\mathrm{r}(l)}\geq P_\mathrm{csp} \\
0 & \mathrm{if} \ \ \Gamma[\overline{P_\mathrm{r}(l)}]\overline{P_\mathrm{r}(l)} < P_\mathrm{csp}.
\end{cases}
\label{equ:eff}
\end{eqnarray}
From this equation, we can define the required power $P_\mathrm{req}$ to activate the sensor node as $P_\mathrm{req}=P_\mathrm{csp}/\Gamma [P_\mathrm{req}]$. Finally, the coverage $C$ of the activation area can be defined as
\begin{eqnarray}
\label{equ:Co}
C = \frac{1}{L}\int_{0}^{L}{A(l) \mathrm{d}l}.
\end{eqnarray}
It is noted that the derived coverage $C$ is a lower bound due to the approximation of Eq.~(\ref{equ:approx}), especially for the case of MPCSD, while it gives the exact values of the theoretical coverage in the case of SP and MP, since the received power is constant against time as shown below. 

Assuming free-space condition, the channel models from the transmitters Tx$\sharp$1 and Tx$\sharp$2 to the sensor node can be respectively expressed as
\begin{eqnarray}
\label{equ:chan}
h_1&=&\frac{\lambda}{4 \pi l} \mathrm{e}^{-j2 \pi \frac{l}{\lambda}} \\
h_2&=&\frac{\lambda}{4 \pi (L-l)} \mathrm{e}^{-j2 \pi \frac{(L-l)}{\lambda}},
\end{eqnarray}
where $\lambda$ is the wavelength of the carrier frequency $f_0$. It is noted that, although different frequencies are used in the two transmitters, its difference $\Delta f$ is very small compared with the carrier frequency $f_0$, such as $\Delta f=1$~kHz and $f_0=916.8$~MHz. Thus, the difference on the wavelength is almost negligible, and we denote it with a single parameter $\lambda$ in this paper. In addition, the transmit signals can be expressed by using the equivalent transmit power $P_\mathrm{t}^\mathrm{e}$ as
\begin{eqnarray}
s_1&=& \sqrt{P_\mathrm{t}^\mathrm{e}} \mathrm{e}^{j(2 \pi f_1 t + \theta_1 )} \\
s_2&=& \sqrt{P_\mathrm{t}^\mathrm{e}} \mathrm{e}^{j(2 \pi f_2 t + \theta_2 )}.
\label{equ:sig}
\end{eqnarray}

From these equations, the received power in the case of single transmitter can be calculated as
\begin{eqnarray}
\label{equ:SP}
%\overline{P_\mathrm{r}^\mathrm{SP1}(l)} &=& \frac{1}{T_\mathrm{d}}\int_{0}^{T_\mathrm{d}}{|h_1 s_1|^2 \mathrm{d}t} = P_\mathrm{t}^\mathrm{e} \left( \frac{\lambda}{4 \pi l}\right)^2 
P_\mathrm{r}^\mathrm{SP1}(l) &=& |h_1 s_1|^2  = P_\mathrm{t}^\mathrm{e} \left( \frac{\lambda}{4 \pi l}\right)^2 
\end{eqnarray}
\begin{eqnarray}
%\overline{P_\mathrm{r}^\mathrm{SP2}(l)} &=& \frac{1}{T_\mathrm{d}}\int_{0}^{T_\mathrm{d}}{|h_2 s_2|^2 \mathrm{d}t} = P_\mathrm{t}^\mathrm{e} \left( \frac{\lambda}{4 \pi (L-l)}\right)^2 .
P_\mathrm{r}^\mathrm{SP2}(l) &=& |h_2 s_2|^2 = P_\mathrm{t}^\mathrm{e} \left( \frac{\lambda}{4 \pi (L-l)}\right)^2 .
\end{eqnarray}
If the both transmitters are used, the received power becomes a function depending on the phase conditions in Eq.~(\ref{equ:chan})-(\ref{equ:sig}) and can be calculated as
\begin{eqnarray}
\label{equ:MP}
%\overline{P_\mathrm{r}(l)} &=& \frac{1}{T_\mathrm{d}}\int_{0}^{T_\mathrm{d}}{|h_1 s_1+h_2 s_2|^2 \mathrm{d}t} \nonumber \\
P_\mathrm{r}(l) &=& |h_1 s_1+h_2 s_2|^2  \nonumber \\
&=& P_\mathrm{t}^\mathrm{e}\left( \frac{\lambda}{4 \pi} \right)^2 \left\{ \frac{1}{l^2} + \frac{1}{(L-l)^2} \right. \nonumber \\ 
%&+& \left.  \frac{2 \overline { \cos \left[2 \pi \Delta ft + \Delta \theta +2 \pi \frac{L}{\lambda} \right]} } {l(L-l)} \right\} . \nonumber \\ 
&+& \left.  \frac{2  \cos \left[2 \pi \Delta ft + \Delta \theta +2 \pi \frac{L-2l}{\lambda} \right] } {l(L-l)} \right\} .  
\end{eqnarray}
In the case of MP ($\Delta f = 0$), the received power becomes
\begin{eqnarray}
&&P_\mathrm{r}^\mathrm{MP}(l) \nonumber \\  &=& P_\mathrm{t}^\mathrm{e}\left( \frac{\lambda}{4 \pi} \right)^2  \left\{ \frac{1}{l^2} + \frac{1}{(L-l)^2} +  \frac{2\cos \left[ \Delta \theta + 2 \pi \frac{L-2l}{\lambda} \right] }{l(L-l)} \right\} . \nonumber \\ 
\end{eqnarray}
In this equation, the third term expresses the interference between the two transmitters. On the other hand, in the case of MPCSD ($\Delta f \neq 0$), the third term in Eq.~(\ref{equ:MP}) can be canceled out by time averaging in the capacitor if $T_\mathrm{f} \ll T_\mathrm{d}$ is satisfied. Therefore, the average received power becomes
\begin{eqnarray}
\label{equ:MPCSD}
\overline{P_\mathrm{r}^\mathrm{MPCSD}(l)}
&=& P_\mathrm{t}^\mathrm{e}\left( \frac{\lambda}{4 \pi} \right)^2 \left\{ \frac{1}{l^2} + \frac{1}{(L-l)^2}\right\} \nonumber \\ 
&=& P_\mathrm{r}^\mathrm{SP1}(l) + P_\mathrm{r}^\mathrm{SP2}(l) .
\end{eqnarray}
From this equation, it is clear that the average received power of MPCSD becomes the summation of the received powers of the SPs. 

\begin{table} [!t]
\renewcommand{\arraystretch}{1.3}
\caption{Estimation parameters.}
\label{tab:Est_para}
\centering
\begin{tabular}{c c || c}
\hline
Transmit power & $P_\mathrm{t}$ & 1 W(= 30~dBm)	\\	\hline				
Tx antenna gain &$G_\mathrm{t}$ & 4 ($\approx$ 6 dBi )\\ \hline	
Rx antenna gain &$G_\mathrm{r}$ & 1.6 ($\approx$ 2.15 dBi ) \\ \hline
Center frequency& $f_\mathrm{1}$ & 916.8~MHz \\ \hline
Frequency offset & $\Delta f$ & 1~kHz \\ \hline
Distance between 2 Txs &$L$ & 6 m\\ \hline
Duty cycle & $T_\mathrm{d}$ & 1~s	\\ \hline
Required power &$P_\mathrm{req}$ & 400 $\mu$W ($\approx -4$~dBm ) \\ \hline  
\end{tabular}
\end{table}
From these theoretical equations, the theoretical coverage in the experiments can be estimated. By using the estimation parameters shown in Table~\ref{tab:Est_para}, the theoretical values of the coverage for SP, MP and MPCSD can respectively become 
\begin{eqnarray}
C^\mathrm{SP}=55.5\%, \ \ C^\mathrm{MP} = 90.9\%, \ \ C^\mathrm{MPCSD} = 100\%. \nonumber
\end{eqnarray}
The estimated results show that the coverage of MPCSD achieves 100\% which is about twice that of SP, while MP cannot achieve the full coverage due to the destructive interference. It is noted that the coverage of MP may be slightly changed by the phase difference $\Delta \theta$, while $\Delta \theta = 0$ is assumed in the estimation. However, the effect of the phase difference can only yield a coverage difference of less than 1\%.

At the last part of this section, we would like to show a design criterion of the distance between two Txs $L$ by using the developed equations. In Eq.~(\ref{equ:MPCSD}), it is clear that the location with minimum received power between two Txs is at $l=L/2$, and that power should be higher than the required power $P_\mathrm{req}$. So that the following condition should be hold as 
\begin{eqnarray}
\overline{P_\mathrm{r}^\mathrm{MPCSD}}(L/2) = P_\mathrm{t}^\mathrm{e} \left( \frac{\lambda}{4 \pi} \right)^2 \left( \frac{2^3}{L^2} \right) \geq P_\mathrm{req}.
\end{eqnarray}
Therefore, the maximum distance to guarantee the 100\% coverage is designed as
\begin{eqnarray}
L_\mathrm{max}^\mathrm{MPCSD} = 2\sqrt{2} \frac{\lambda}{4 \pi} \sqrt{\frac{P_\mathrm{t}^\mathrm{e}}{P_\mathrm{req}}}.
\end{eqnarray}
Since the maximum distance $L_\mathrm{max}^\mathrm{SP}$ for SP is simply calculated as 
\begin{eqnarray}
L_\mathrm{max}^\mathrm{SP} = \frac{\lambda}{4 \pi} \sqrt{\frac{P_\mathrm{t}^\mathrm{e}}{P_\mathrm{req}}},
\end{eqnarray}
$L_\mathrm{max}^\mathrm{MPCSD}$ is $2\sqrt{2}$ times longer than that of SP. Thus, MPCSD works not only for the interference mitigation between two Txs, but also for the extension of the distance of two Txs $L$.

\section{Development of low-energy battery-less sensor node}
\label{sec:BLS}
This section develops real battery-less sensor node composed of off-the-shelf devices, provides the design criteria and measures power consumption of the sensor node and the RF/DC conversion efficiency in order to understand the property of realistic consumed power of sensor nodes and the benefits of the intermittent operation. In addition, the parameters for the experimental investigation to be presented in the next section, i.e. RF/DC conversion efficiency and required power of the activation, will be provided by the measurement results of this section.

\subsection{Prototype Hardware}
\begin{figure}[!t]
    \centering
    \includegraphics[width=4in]{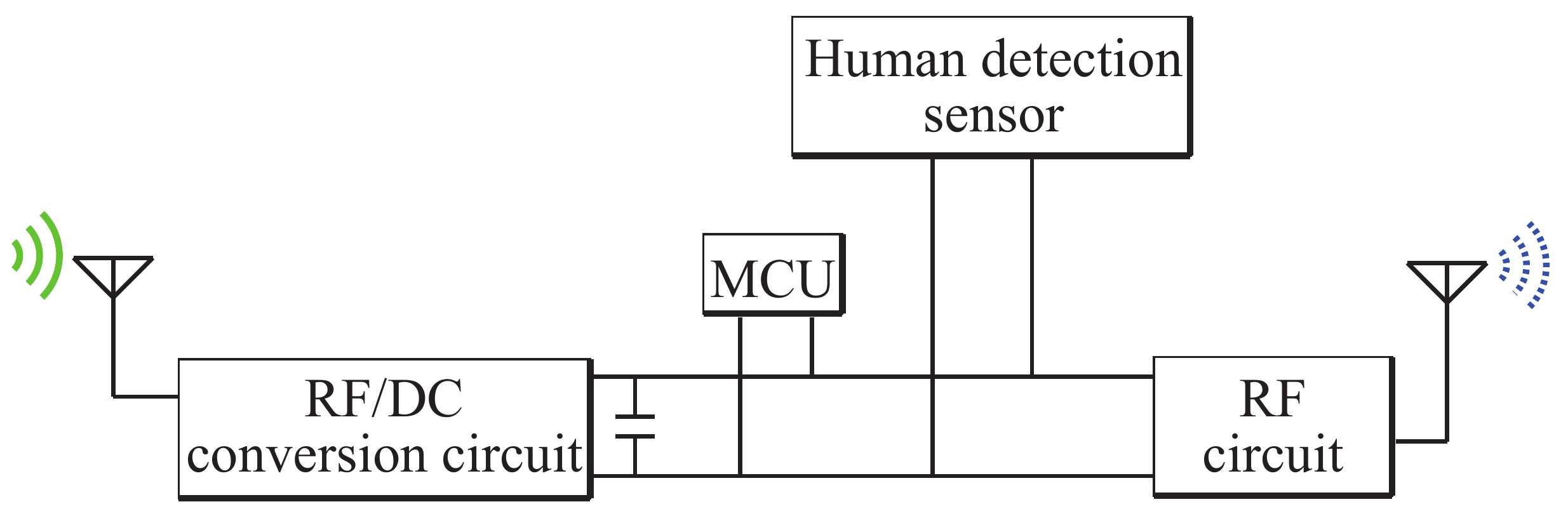}
    \caption{Architecture of battery-less sensor.}
    \label{fig:sensor_com}
\end{figure}

\begin{figure}[!t]
    \centering
    \includegraphics[width=4.5in]{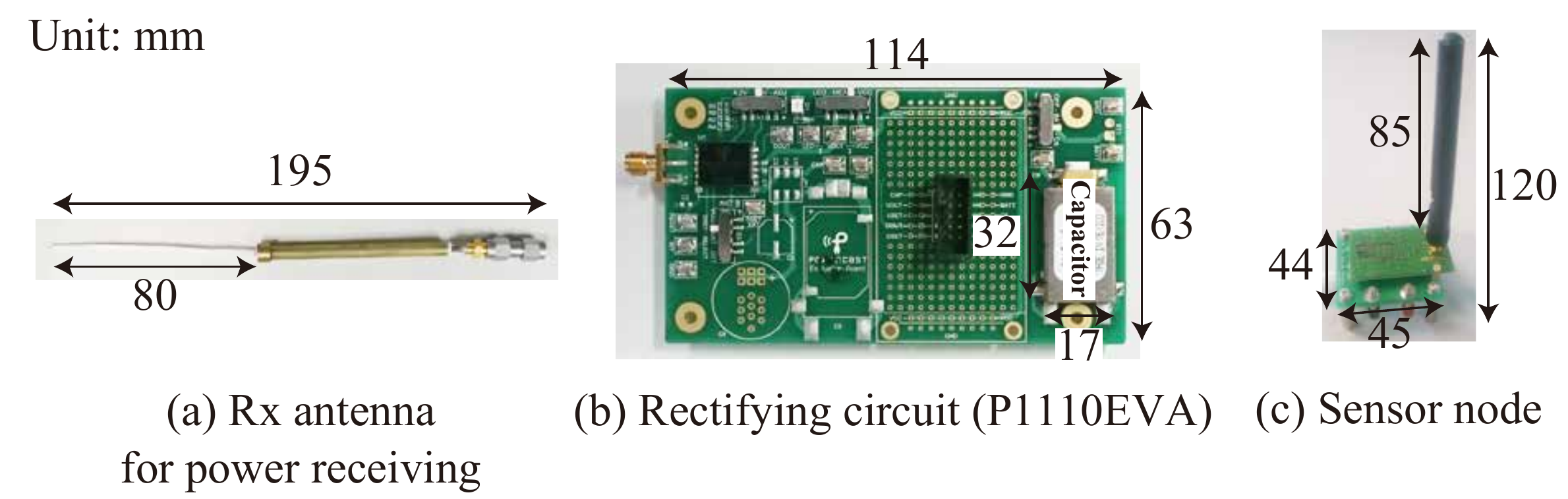}
    \caption{Photos of components of battery-less sensor.}
    \label{fig:com_pic}
\end{figure}

Table~\ref{tab:sensor_com}, Figs.~\ref{fig:sensor_com} and \ref{fig:com_pic} describe the components of the prototype hardware. The battery-less sensor node mainly consists of a power receiving antenna in Fig.~\ref{fig:com_pic}~(a), a rectifying (RF/DC conversion) circuit  including a capacitor of 50~mF in Fig.~\ref{fig:com_pic}~(b) \cite{Powercast}, and a sensor node in Fig.~\ref{fig:com_pic}~(c) including a MCU (Micro Controller Unit) and  an IR (InfraRed) human detection sensor. The RF energy is received by the antenna and is converted to DC required for RF module by the rectifying circuit. The MCU in the RF module manages the data transmission, while reducing the power consumption whose details are given in the following subsections.
 
\begin{table} [!t]
\renewcommand{\arraystretch}{1.3}
\caption{Components of Battery-less Sensor.}
\label{tab:sensor_com}
\centering
\begin{tabular}{c || c}
\hline
RF module & Custom made (TESSERA Technology)	\\					
RF device & ADF7023-J (Analog Devices) \\ 
MCU & R5F100GJ (Renesas Electronics)\\
Human detection sensor & EKMB1101111 (Panasonic)\\
RF/DC conversion circuit & P1110EVB (Powercast) \\
\hline  
\end{tabular}
\end{table}

\subsection{Low-energy design}
\begin{figure}[t]
\centering
\includegraphics[width=4in]{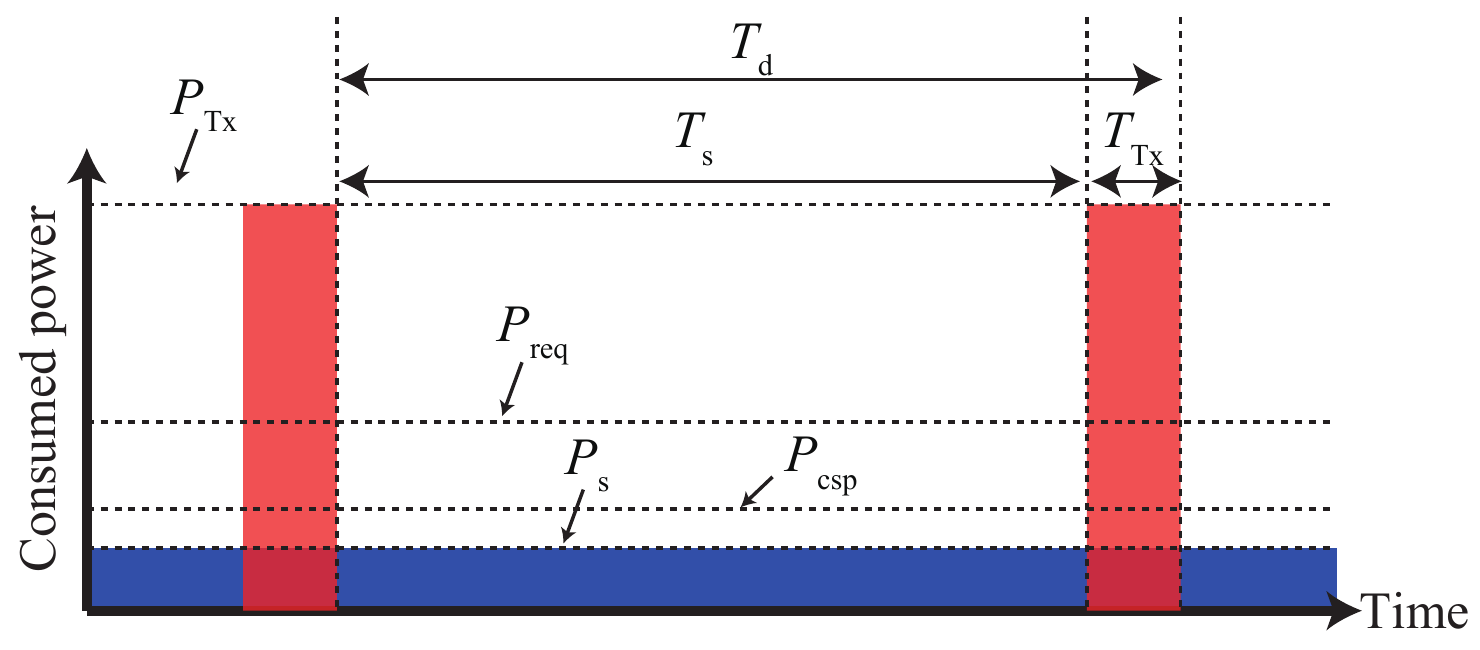}
\caption{Concept of intermittent operation.}
\label{fig:IO}
\end{figure}

To extend the coverage, where sensor nodes can be activated, the consumed power of the sensor node should be as low as possible. To reduce the power consumption, an intermittent operation of data transmission can be introduced and operated by MCU. Since the consumed power of sensor is generally much less than that of RF process, only RF process is performed periodically. On the other hand, because sensor often has initialization time for e.g. calibration, the sensor itself is always active. In addition, a capacitor equipped in the sensor node also plays the role of a rechargeable battery which recharges the surplus energy in the sleep mode so that the sensor nodes can be stably activated even when the consumed power of data transmission is larger than that of the output power of the rectifying circuit. Here, the consumed power of sensor node is described as
\begin{eqnarray}
P_{\mathrm{csp}}\left( t \right) = \begin{cases}
P_{\mathrm{s}}&(0 <  t \leq T_{\mathrm{s}} ) \\
P_{\mathrm{Tx}}&(T_{\mathrm{s}}< t \leq T_{\mathrm{d}}), 
\end{cases}
\label{equ:cons}
\end{eqnarray}
where $T_{\mathrm{s}}(=T_{\mathrm{d}}-T_{\mathrm{Tx}})$ and $T_{\mathrm{Tx}}$ are the duration of the sleep and Tx modes respectively, $T_{\mathrm{d}}$ is the duty cycle, $P_{\mathrm{s}}$ is the consumed power in the sleep mode, and $P_{\mathrm{Tx}}$ is the consumed power in the Tx mode as shown in Fig. \ref{fig:IO}.  From Eq. (\ref{equ:cons}), the average consumed power becomes
\begin{eqnarray}
P_{\mathrm{csp}} = \frac{1}{T_{\mathrm{d}}}\left( \int_{0}^{T_{\mathrm{s}}} P_{\mathrm{s}} dt + \int_{T_{\mathrm{s}}}^{T_{\mathrm{d}}} P_{\mathrm{Tx}}  dt \right).
\label{equ:csp}
\end{eqnarray}
Based on this equation, the average consumed power $P_{\mathrm{csp}}$ can be reduced by increasing $T_{\mathrm{d}}$ or decreasing $P_{\mathrm{Tx}}$ and $P_{\mathrm{s}}$.  
Because $T_\mathrm{d}$ is determined by the application of WSN, the other two factors should be as low as possible.

\begin{figure}[t]
    \centering
    \includegraphics[width=1.5in]{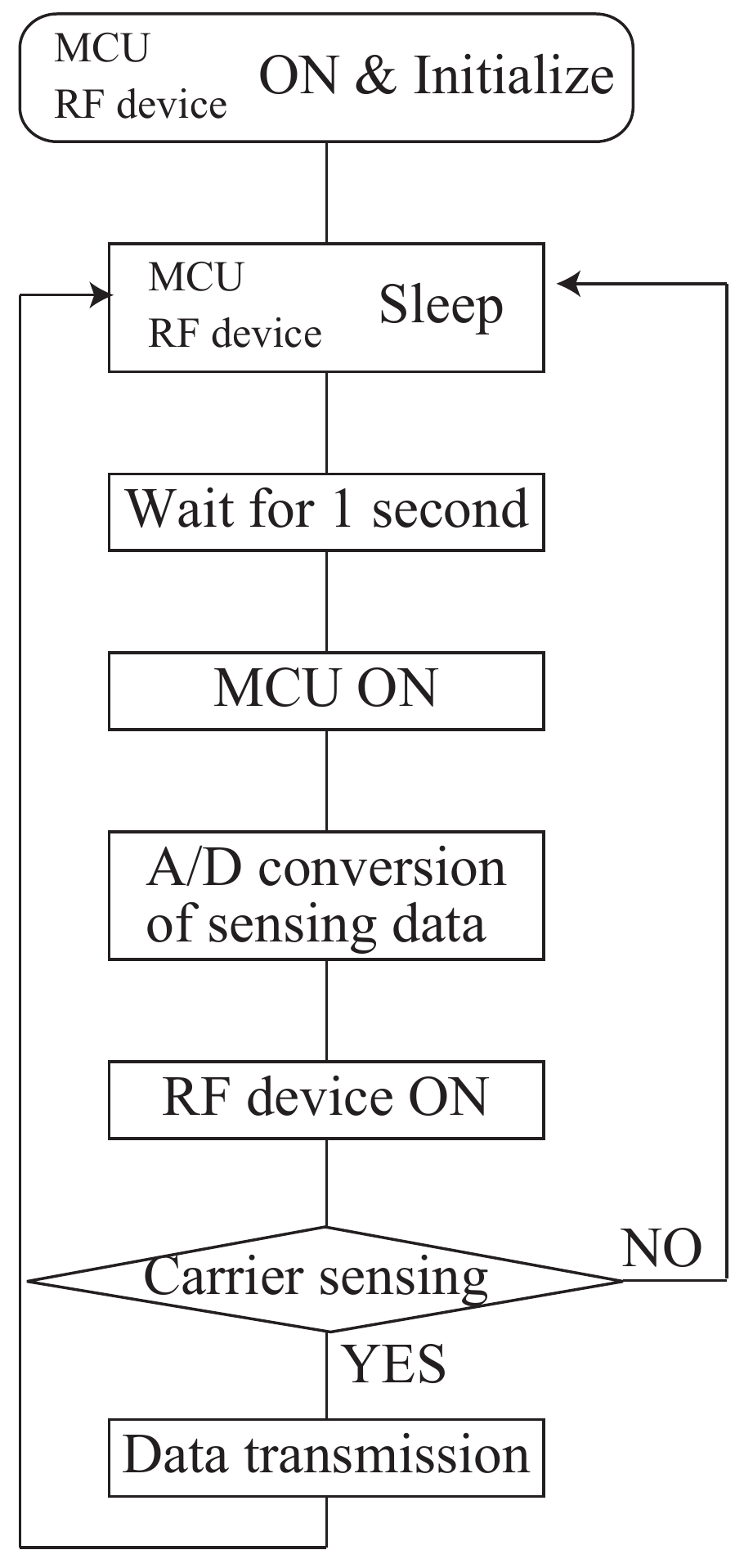}
    \caption{Activation flow of sensor node.}
    \label{fig:FL}
\end{figure}

Figure~\ref{fig:FL} shows the operation flow of sensor node to reduce consumed power. In the sleep mode, only the sensor is activated because the human detection sensor requires a long initialization time of 15~s while low power consumption is realized by sleeping the MCU and RF device. After the MCU is activated by itself in the Tx mode, the MCU performs data transmission with carrier sensing. It is noted that even if the data transmission cannot be performed due to collision avoidance, the MCU returns to the sleep mode because the power consumption of the carrier sensing is almost the same as that of data transmission. In this paper, the duty cycle is set to 1~s since we assume an application of human detection.

In order to manage the intermittent operation while reducing the size of sensor node, the value of capacitor should be designed carefully. In the capacitor, the stored energy should be higher than the required energy of Tx mode. When the received power is the same with required power $P_\mathrm{r} = P_\mathrm{req}$, the following condition should be hold at the capacitor to activate the sensor node,
\begin{eqnarray}
\frac{\mathfrak{C}}{2} \left[ V_\mathrm{c}^2 (T_\mathrm{s}) - V_\mathrm{c}^2 (T_\mathrm{d}) \right] &\geq& (P_\mathrm{Tx} - P_\mathrm{req}\Gamma[P_\mathrm{req}])T_\mathrm{Tx} \nonumber \\ &=& (P_\mathrm{Tx} - P_\mathrm{csp})T_\mathrm{Tx},
\end{eqnarray} 
where $\mathfrak{C}$ is the value of the capacitance and $V_\mathrm{c}(t)$ is the voltage at the capacitor at time $t$. Therefore, the minimum required capacitance $\mathfrak{C}_\mathrm{min}$ can be calculated as
\begin{eqnarray}
\mathfrak{C}_\mathrm{min} = 2 T_\mathrm{Tx} \frac{P_\mathrm{Tx} - P_\mathrm{csp}}{ V_\mathrm{c}^2 (T_\mathrm{s}) - V_\mathrm{c}^2 (T_\mathrm{d}) } . 
\end{eqnarray} 
For the developed sensor node, the minimum value is calculated as about 600 $\mu$F by substituting parameters measured in the next section
($T_\mathrm{Tx} = 10$~ms, $P_\mathrm{Tx}=13.8$~mW, $P_\mathrm{csp}=142$~$\mu$W) and the typical and minimum operating voltage of the sensor node ($V_\mathrm{c}(T_\mathrm{s})=$2.3~V, $V_\mathrm{c}(T_\mathrm{d})=$2.2~V) into the equation. However, our developed sensor node employs a 50~mF capacitor since we use a pre-implemented capacitor on the RF/DC conversion board.

\begin{figure}[thb]
    \centering
    \includegraphics[width=3in]{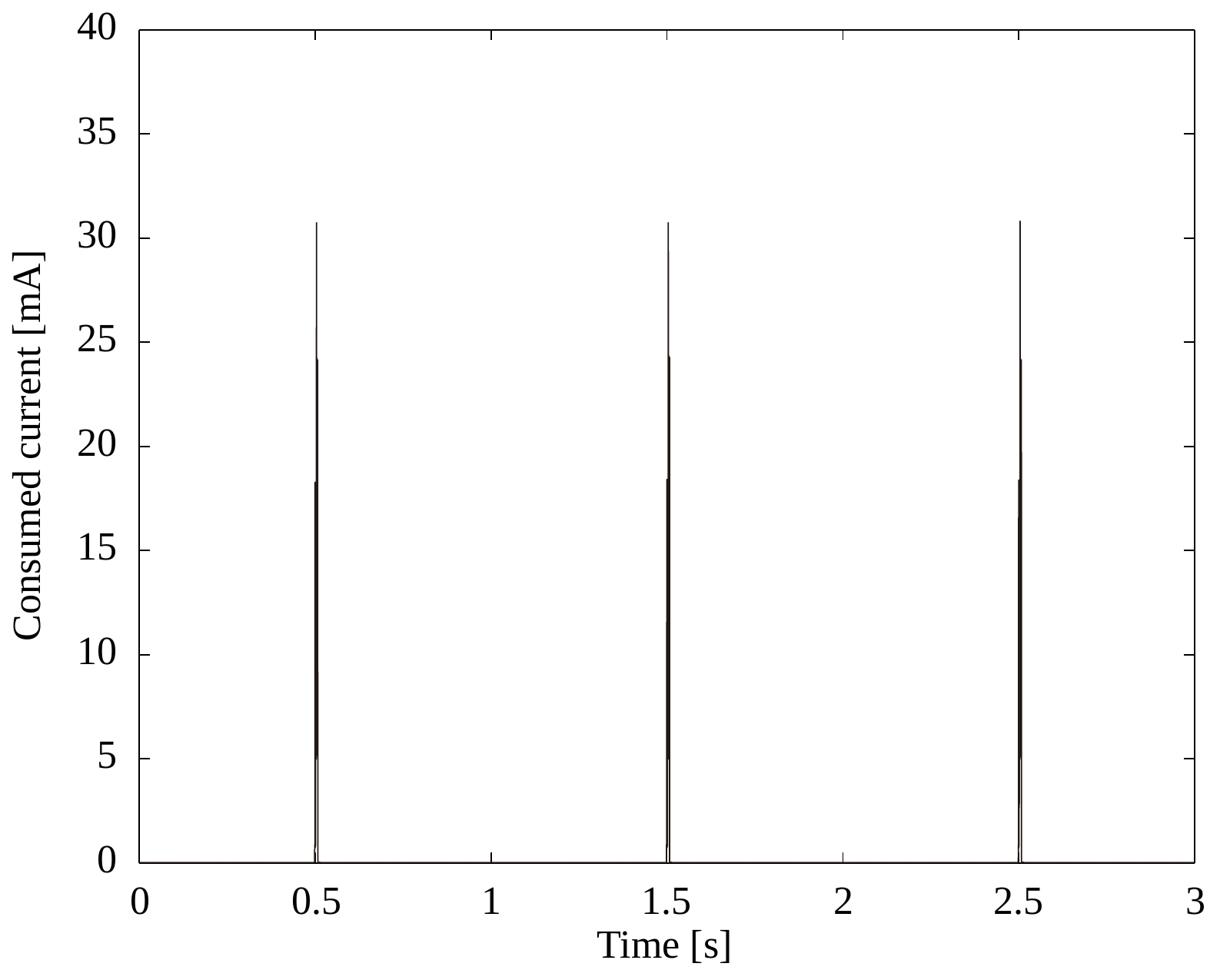}
    \caption{Consumed current of battery-less sensor node in 3 s.}
    \label{fig:RTTx}
\end{figure}

\begin{figure}[!t]
    \centering
    \includegraphics[width=5in]{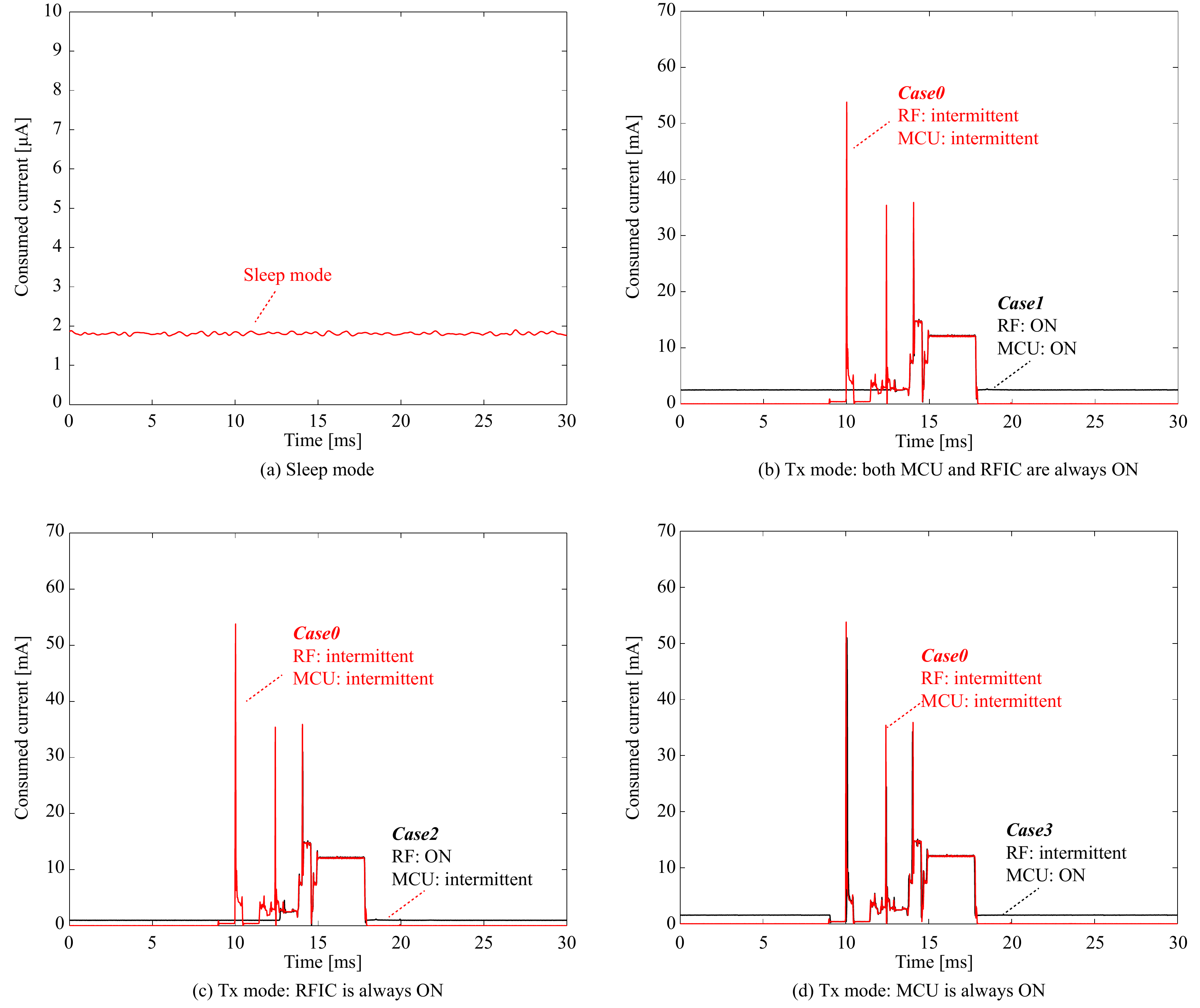}
    \caption{Consumed current of battery-less sensor node in Tx mode.}
    \label{fig:RTCC}
\end{figure}
\begin{figure}[!t]
    \centering
    \includegraphics[width=5in]{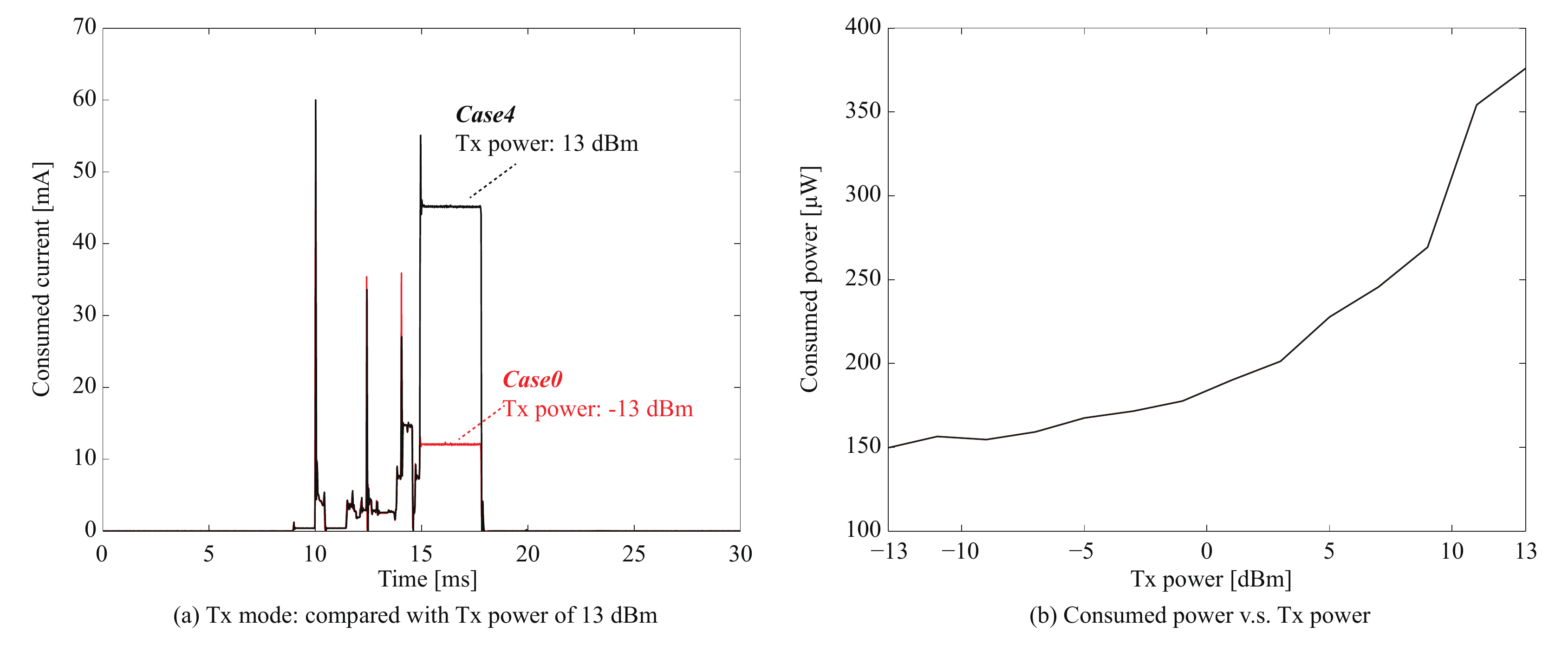}
    \caption{Relationship between consumed power and Tx power of data transmission.}
    \label{fig:High_power}
\end{figure}

\subsection{Measurement on power consumption}
\label{sec:meas_cons}
Figure~\ref{fig:RTTx} shows the measured consumed current of the sensor node in 3~s. As shown in Fig. \ref{fig:RTTx}, the data transmission is performed every 1~s. Figure \ref{fig:RTCC}(a) shows the consumed current of the sleep mode in 30~ms. In the sleep mode, the consumed power is 4.23~$\mu$W. Figures~\ref{fig:RTCC}(b)-(d) show the consumed current of the sensor node in Tx mode. In these figures, the red lines ({\bf \emph{case0}}) show the proposed operation in the case when both RF circuit and MCU employ sleep mode. The black lines in Figs. \ref{fig:RTCC}(b)-(d), show the consumed current in the case when both RF circuit and MCU do not employ the sleep mode ({\bf \emph{case1}}), when only MCU employs the sleep mode ({\bf \emph{case2}}) and when only RF circuit employs the sleep mode ({\bf \emph{case3}}) respectively. In these configurations, the average consumed powers are 142~$\mu$W ({\bf \emph{case0}}), 5.89~mW ({\bf \emph{case1}}), 2.35~mW ({\bf \emph{case2}}), and 3.72~mW ({\bf \emph{case3}}), respectively. As shown in the figures, Tx mode lasts for about 10~ms so that the duty factor is $1/100$. Under the duty factor, the intermittent operation can reduce the consumed power to less than that of $1/10$. It is noted that the voltage is set to a constant value of 2.3~V in the current measurement.

In Fig.~\ref{fig:High_power}(a), the red line shows the consumed current in the case when the Tx power of data transmission is set to $-$13~dBm ({\bf \emph{case0}}) while the black line shows that with 13~dBm ({\bf \emph{case4}}) which is the hardware limitation of maximum Tx power. As shown in Fig.~\ref{fig:High_power}, the data transmission is performed at the time around 15~ms and lasts for 4~ms. In addition, the consumed current of data transmission in {\bf \emph{case0}} can be reduced from about 45~mA to about 10~mA. In addition, Fig.~\ref{fig:High_power}(b) shows the consumed power of sensor node against the Tx power. The curve is similar to the data sheet of power amplifier of RF circuit \cite{AD}. By using sensor node of {\bf \emph{case0}} and the commercial product of rectifying circuit, the required power of battery-less sensor node becomes about 400~$\mu$W ($P_\mathrm{req} \approx  -4$ dBm) which is measured by employing a signal generator with a variable input power into the RF/DC circuit.

\subsection{Measurement on RF/DC conversion efficiency}
In order to understand the property of the RF/DC conversion efficiency, we setup a preliminary experiment to measure the efficiency by inputting signal with variable power generated by signal generator into the RF/DC conversion circuit and observing the voltage of the capacitor. Here, the output of RF/DC circuit is connected to a sensor node configured in sleep mode, knowing that in actual situation an operating sensor node will spend most of its duty cycle in sleep mode.  Because the difference between the output power of the RF/DC circuit and the consumed power in sleep mode should be stored in the capacitor, the following equation should be hold.
\begin{eqnarray}
(P_\mathrm{in} \Gamma[P_\mathrm{in}] - P_{\mathrm{s}})T = \frac{\mathfrak{C}}{2}(V^2_\mathrm{c}(T) - V^2_\mathrm{c}(0) ), 
\end{eqnarray}
where $P_\mathrm{in}$ is the input power and $T$ is the measurement time. From this equation, the RF/DC conversion efficiency can be derived as
\begin{eqnarray}
\Gamma[P_\mathrm{in}]  = \left( \frac{\mathfrak{C}}{2T}  (V^2_\mathrm{c}(T) - V^2_\mathrm{c}(0) )  +  P_{\mathrm{s}} \right) / P_\mathrm{in}.
\end{eqnarray}
Here, the capacitance $\mathfrak{C}$ of 50~mF is the same with the developed sensor node and the measurement time $T$ of 20~s is the same as that of the experiment presented in the next section. 

\begin{figure}[t]
    \centering
    \includegraphics[width=3in]{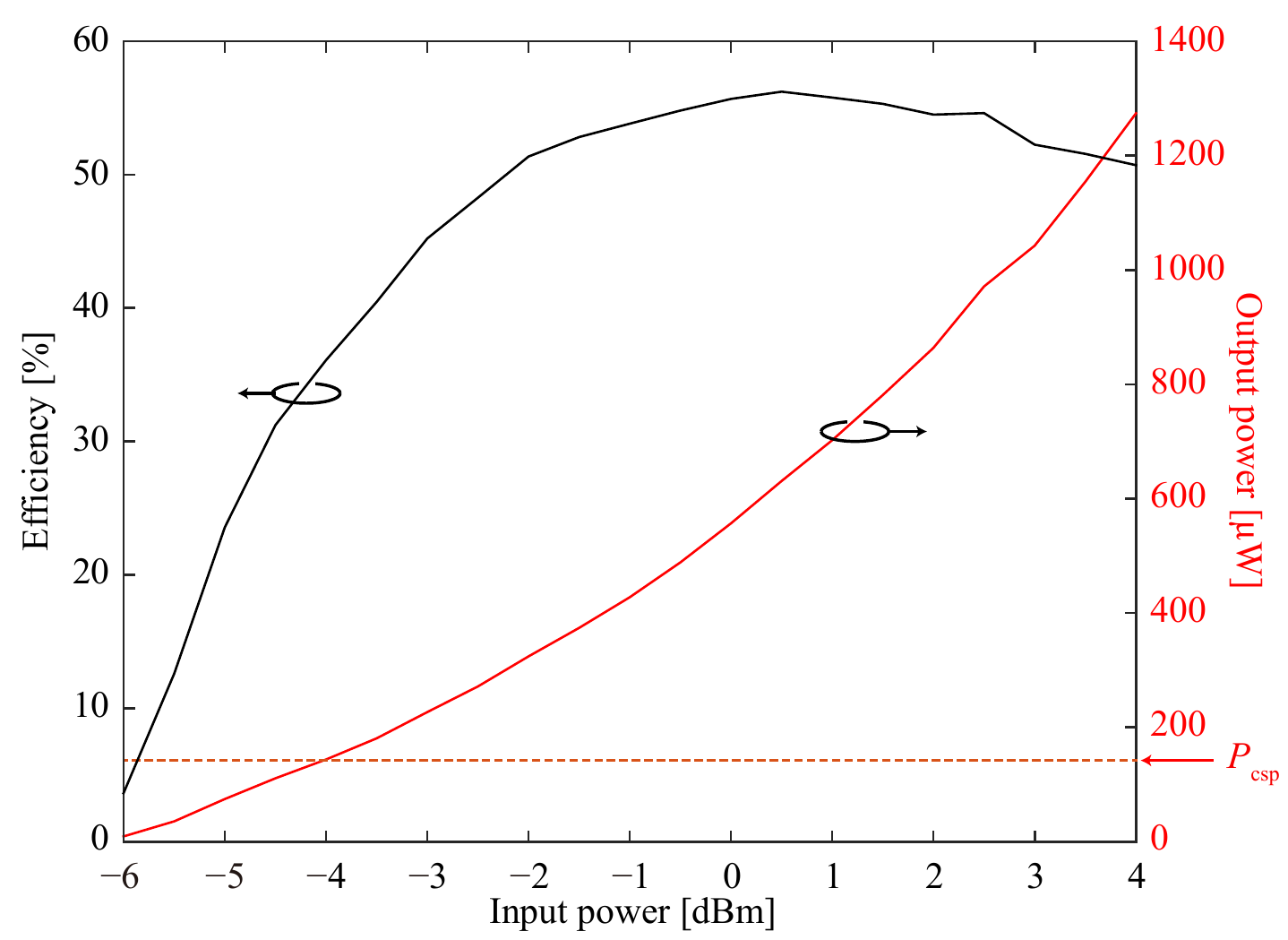}
    \caption{Measurement of RF/DC conversion efficiency.}
    \label{fig:Eff_meas}
\end{figure}

Figure~\ref{fig:Eff_meas} shows the result of the preliminary measurement. As shown in the figure, the efficiency has non-linearity against input power. However, the output power is monotonically increases against the input power, so that the required power can be defined as the threshold power to activate the sensor node. It is noted that when $P_\mathrm{in}=P_\mathrm{req}=-4$~dBm, the output power is almost the same with the average consumed power of the sensor node measured in Sect.~\ref{sec:meas_cons}. In addition, from this result, we have confirmed that the assumption in Sect.~\ref{sec:Est} $\partial^2 \left( \Gamma[P_\mathrm{r}(l)] P_\mathrm{r}(l) \right)/\partial {P_\mathrm{r}(l)}^2~\ge~0$ holds in the region where the output power is less than 1~mW (0~dBm). 

\section{Indoor experiment}
\label{sec:Ex}
\subsection{Experiment method}
\begin{figure}[!t]
    \centering
    \includegraphics[width=4in]{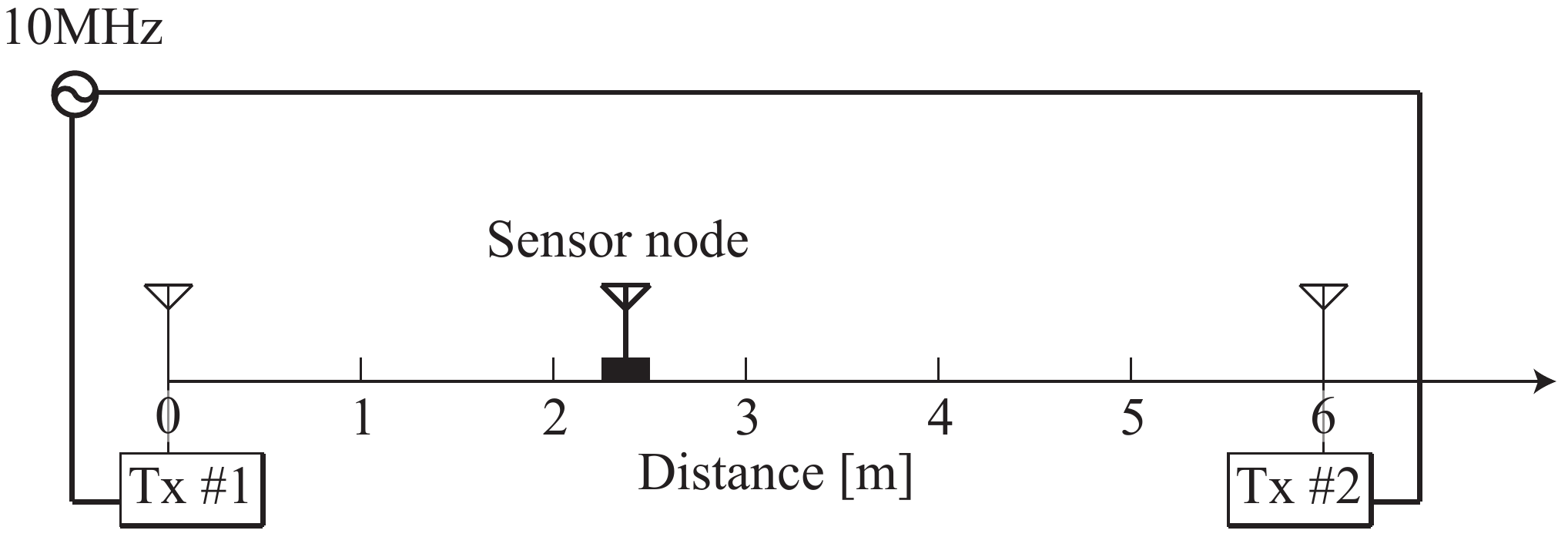}
    \caption{Experimental environment.}
    \label{fig:Ex}
\end{figure}

\begin{figure}[!t]
    \centering
    \includegraphics[width=4in]{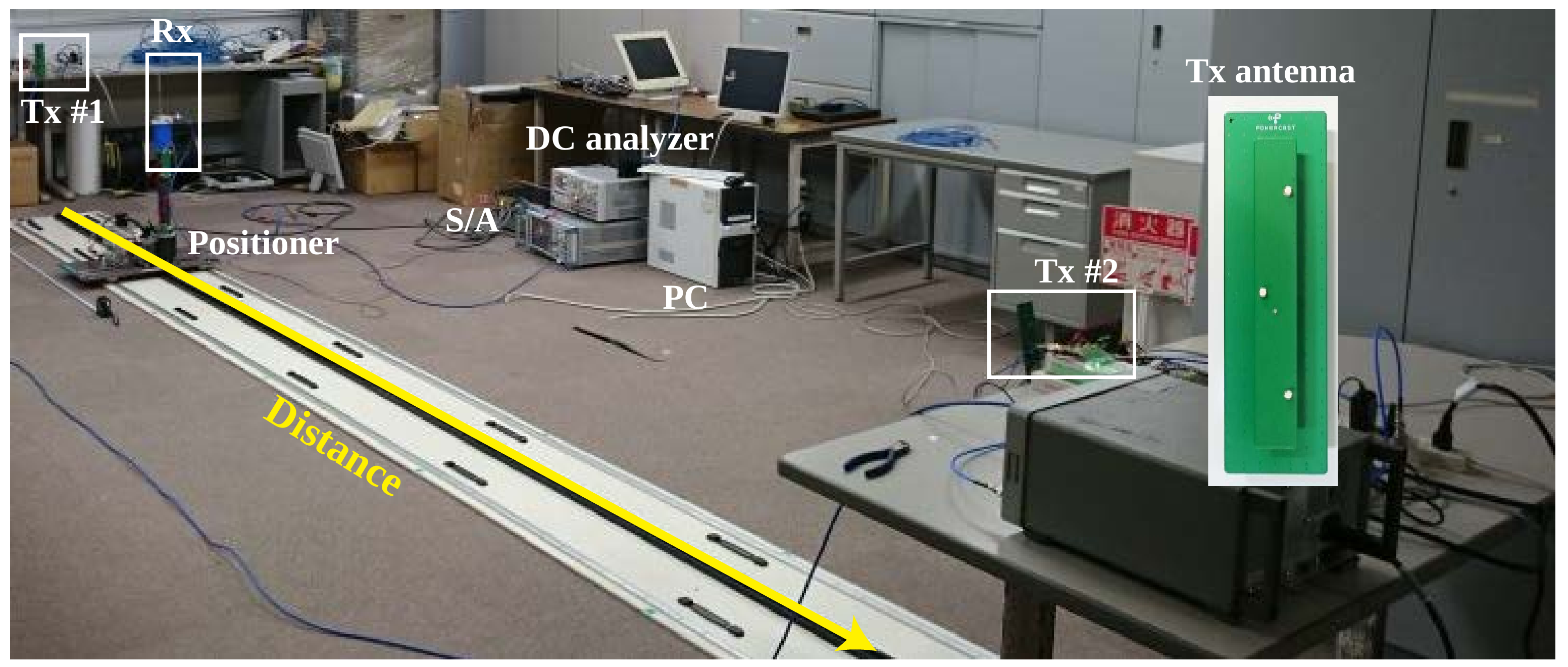}
    \caption{Photo of experimental environment.}
    \label{fig:Ex_pic}
\end{figure}

We conduct indoor experiments to verify the effectiveness of MPCSD with the prototype hardware. Table \ref{tab:Ex}, Figs. \ref{fig:Ex} and \ref{fig:Ex_pic} show the experimental parameters and the experimental environment respectively. At the Tx, EIRP is set to 36~dBm and the center frequency is set to the 916.8~MHz at which transmission without carrier sensing is allowed. The carrier offset is set to 1~kHz to generate the cycle of artificial power fluctuation of 1~ms which is much faster than 1~s of the duty cycle. 
The measurement is performed along the straight line between Tx $\sharp1$ and $\sharp2$ as shown in Fig.~\ref{fig:Ex}. Both the Tx and Rx antennas are equipped at the same height of 0.82~m which is the same as the height of desk because we assume the application of detecting human existence at their working desks. Due to the multipath effect, the propagation pattern would be changed by employing a different antenna height. However, the MPCSD can mitigate such multipath effect as analyzed in \cite{Paper1}. Therefore, the experiments with different antenna heights are out of scope of this paper. It is noted that deployment of Tx antennas on ceiling as shown in Fig.~\ref{fig:MPC} is effective to supply energy as reported in \cite{Smart}\cite{APMC}. However, such configuration increases the complexity of the experimental setup because of the requirement for a careful consideration on antenna directivity. For that reason, attaching Tx antennas on the ceiling is also out of scope of this paper.
In the cases of MP and MPCSD, two transmitters are synchronized by the 10~MHz oscillator.
Other parameters related to data transmission are shown in Tab.~\ref{tab:Ex}, where the data rate and modulation conform to IEEE 802.15.4 \cite{802}\cite{Tessera}.

\begin{figure}[!t]
    \centering
    \includegraphics[width=4in]{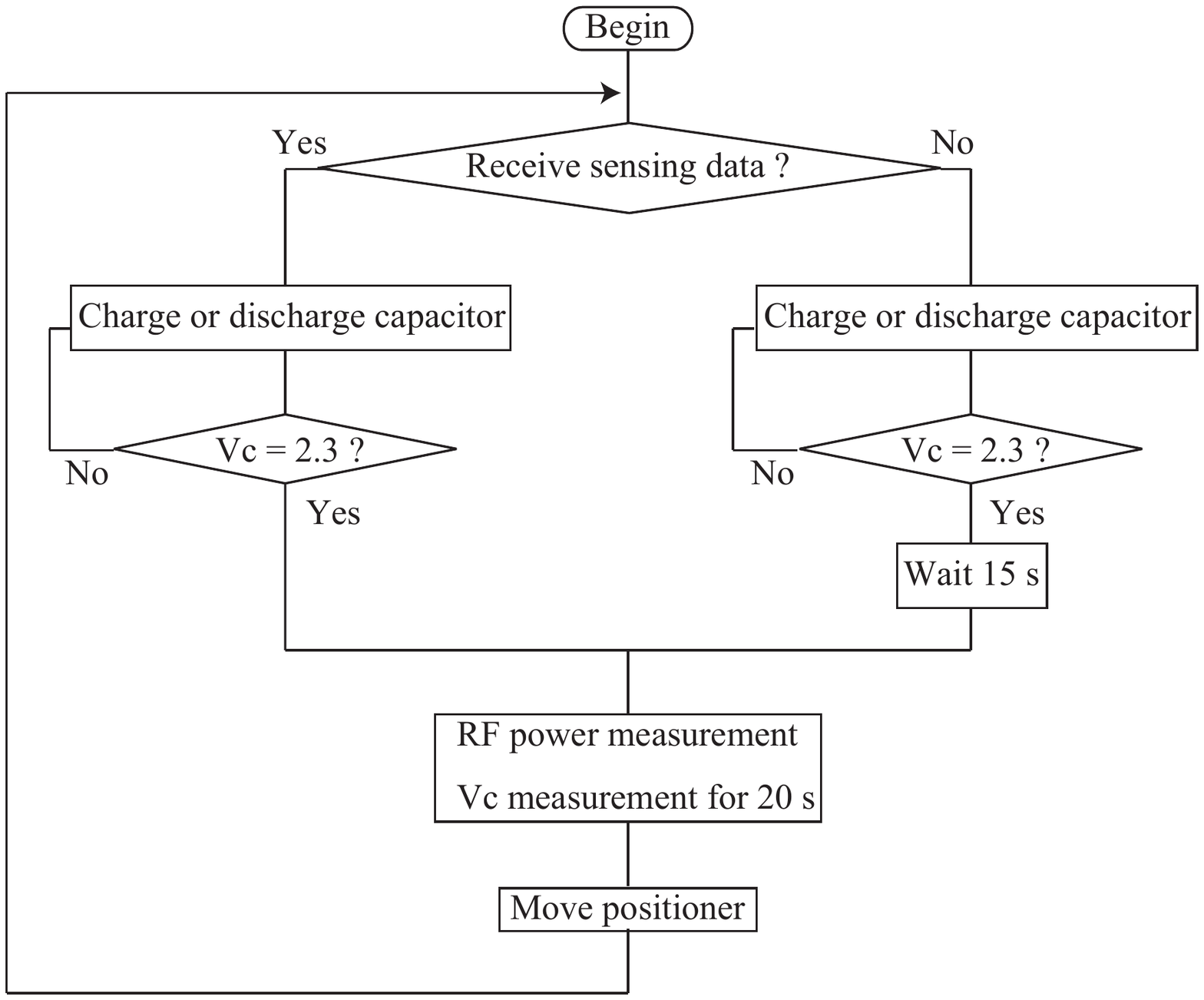}
    \caption{Experimental flow.}
    \label{fig:Ex_flow}
\end{figure}

Figure~\ref{fig:Ex_flow} shows the flowchart of the experiment. To keep the fairness for all measurement points, a capacitor is charged/discharged to be 2.3~V before the measurement. In the experiment, the definition of the activation is judged by whether the measured voltage is increased (active) or decreased (inactive) in 20~s. Since IR sensor takes several seconds for initial activation, at the first step of the measurements, a PC with data receiver checks whether sensor is activated or not. If sensor node is not activated, the measurement is stopped in 15~s until finishing the initialization of IR sensor. After the voltage setup, RF received power is measured and the voltage of the capacitor is measured in 20~s. 

\begin{figure}[!t]
    \centering
    \includegraphics[width=4in]{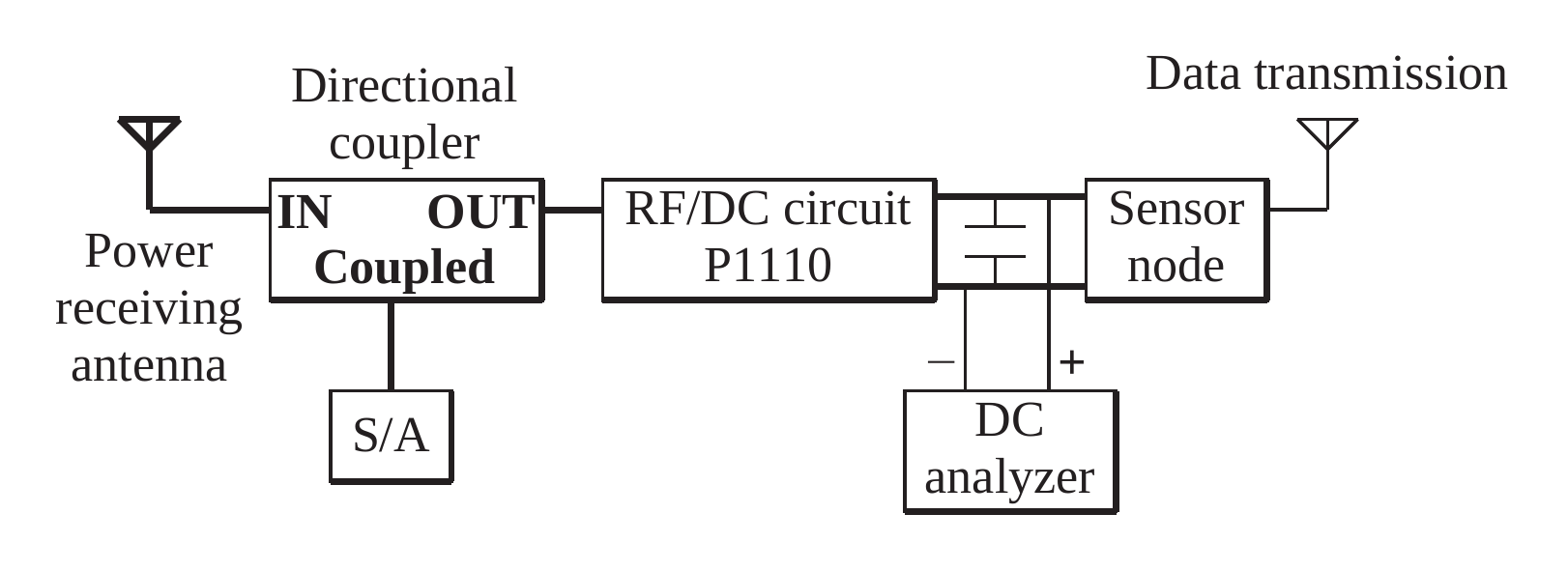}
    \caption{Receiver network.}
    \label{fig:Rx_net}
\end{figure}

Figure~\ref{fig:Rx_net} shows the receiver network. RF signal is received at the power receiving antenna and the signal is transfered to the RF/DC conversion circuit through a 20~dB directional coupler whose coupled line is connected to a S/A (Spectral Analyzer) measuring the received power. The output of RF/DC conversion circuit is connected to the sensor node and a DC analyzer to measure the voltage of the capacitor. In general, the antenna port of the rectifying circuit connected to the output of the coupler is not designed for 50$\Omega$ network, while the power receiving antenna, directional coupler and S/A is designed for 50$\Omega$ network. Therefore, the impedance mismatch could occur between 50$\Omega$ and non-50$\Omega$ networks and results in the discrepancy between the real received power and the measurement value of S/A. In addition, the impedance of the non-50$\Omega$ network varies against the input power of the rectifying circuit due to the diode's non-linearity and against the voltage of the capacitor. \cite{Zoya1}\cite{Zoya2} have shown the degradation of the efficiency due to the impedance mismatch and the non-linearity. In this experiment, because the rectifying circuit is designed to connect 50$\Omega$ antenna, the mismatch effect can be mitigated. In addition, the large value of the capacitor of 50~mF and the coupler isolation of 20~dB suppress the mismatch effect. By a test measurement, in which the rectifying circuit is directly input from a signal generator, it is confirmed that the measurement errors at the S/A are less than 1~dB.

\begin{table} [!t]
\renewcommand{\arraystretch}{1.3}
\caption{Experimental parameters.}
\label{tab:Ex}
\centering
\begin{tabular}{c || c | c}
& Parameter & Value \\
\hline
\hline			
& EIRP & 36 dBm \\ 
Tx& Center frequency of energy transmission & 916.8 MHz\\
& Carrier shift & 1 kHz\\
\hline
& Measurement point interval & 3 cm$\approx \frac{\lambda}{10}$ \\ 
& Center frequency of data transmission & 927.6 MHz \\
Rx & Data rate & 100 kbps \\
&Modulation & Gaussian FSK \\ 
\end{tabular}
\end{table}

\begin{figure}[!t]
    \centering
    \includegraphics[width=5in]{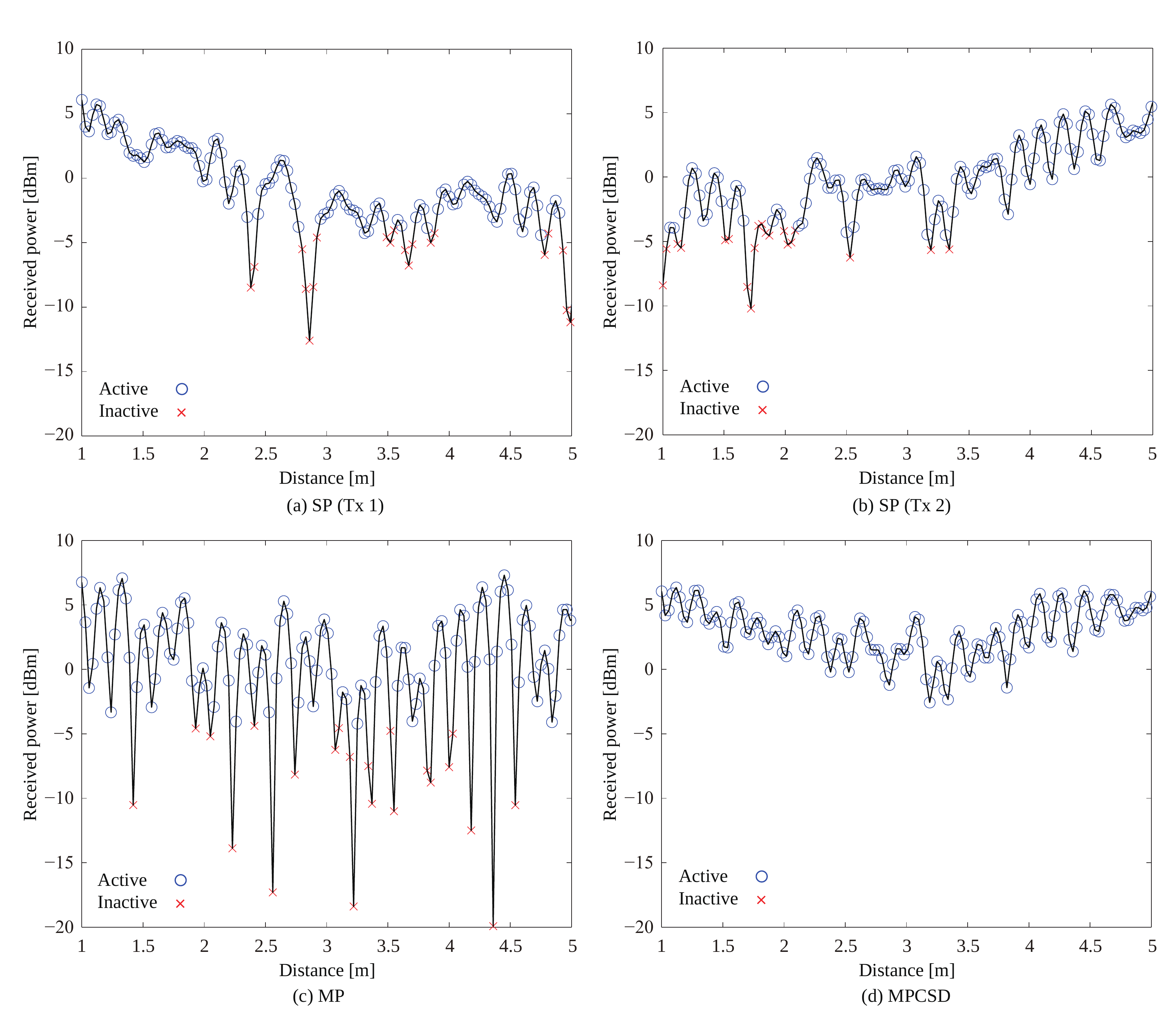}
    \caption{Experimental results.}
    \label{fig:EXR}
\end{figure}

\subsection{Experimental results}
Figures \ref{fig:EXR}(a)-(d) show the experimental results of power distribution and sensor activation. The symbols `O' and `X' correspond to the active and inactive status of the battery-less sensor node respectively. In the case of SP, when the sensor node is far from the corresponding Tx antenna, the received power attenuates and the number of active sensor nodes decreases in proportion to the distance between Tx and sensors as shown in Figs. \ref{fig:EXR}(a) and (b). In these figures, the received power is fluctuated by multipath such that some sensor nodes, which are close to Tx i.e. around 2.3~m in Fig.~\ref{fig:EXR}(a), cannot be activated due to destructive standing-wave, while some sensors, which are far from Tx i.e. around 4.5~m in Fig.~\ref{fig:EXR}(a), can still be activated owing to constructive standing-wave. In the case of MP, several deadspots exist due to destructive interference between multiple wave sources, especially at the central area between the two Txs as shown in Fig.~\ref{fig:EXR}(c). On the contrary, in MPCSD, the degraded received power at deadspots are remarkably improved and the sensor node can be activated at all locations as shown in Fig.~\ref{fig:EXR}(d).  

The activation threshold is about $-4$~dBm which is as expected from $P_{\mathrm{req}}$. However, there are discrepancy points where the sensor node can be activated by a received power even below $-4$~dBm, or where the sensor node cannot be activated by a received power even above $-4$~dBm. This is because the finite measurement time could result in a discrepancy of the activation status because, when the received power is almost the same as the required power, the voltage difference at the capacitor would fluctuate between plus and minus.   The coverage of SP from Tx$\sharp1$ and Tx$\sharp2$ are respectively limited by 84.4\% and 85.2\%. In the case of MP, the coverage is still limited by 83.7\%. On the other hand, the coverage of MPCSD can achieve a full coverage to activate the battery-less sensor node. 

Here, because the measurement cannot be performed in the range of 0~m to 6~m due to the absolute maximum rating of RF/DC conversion circuit, the theoretical coverage calculated in Sect.~\ref{sec:MPC} is recalculated for the reduced range of 1~m to 5~m. The measured results and estimated results are shown in Table~\ref{tab:Co}. In SP, the measured coverage is higher than the theoretical value because of the effect of constructive standing-wave created by multipath. On the other hand, in MP, the measured coverage is almost the same as the theoretical value, so that the effect of multipath is not very dominant in MP. In addition, the measured coverage of MPCSD achieves 100\% as expected.  

\begin{figure}[!t]
    \centering
    \includegraphics[width=3in]{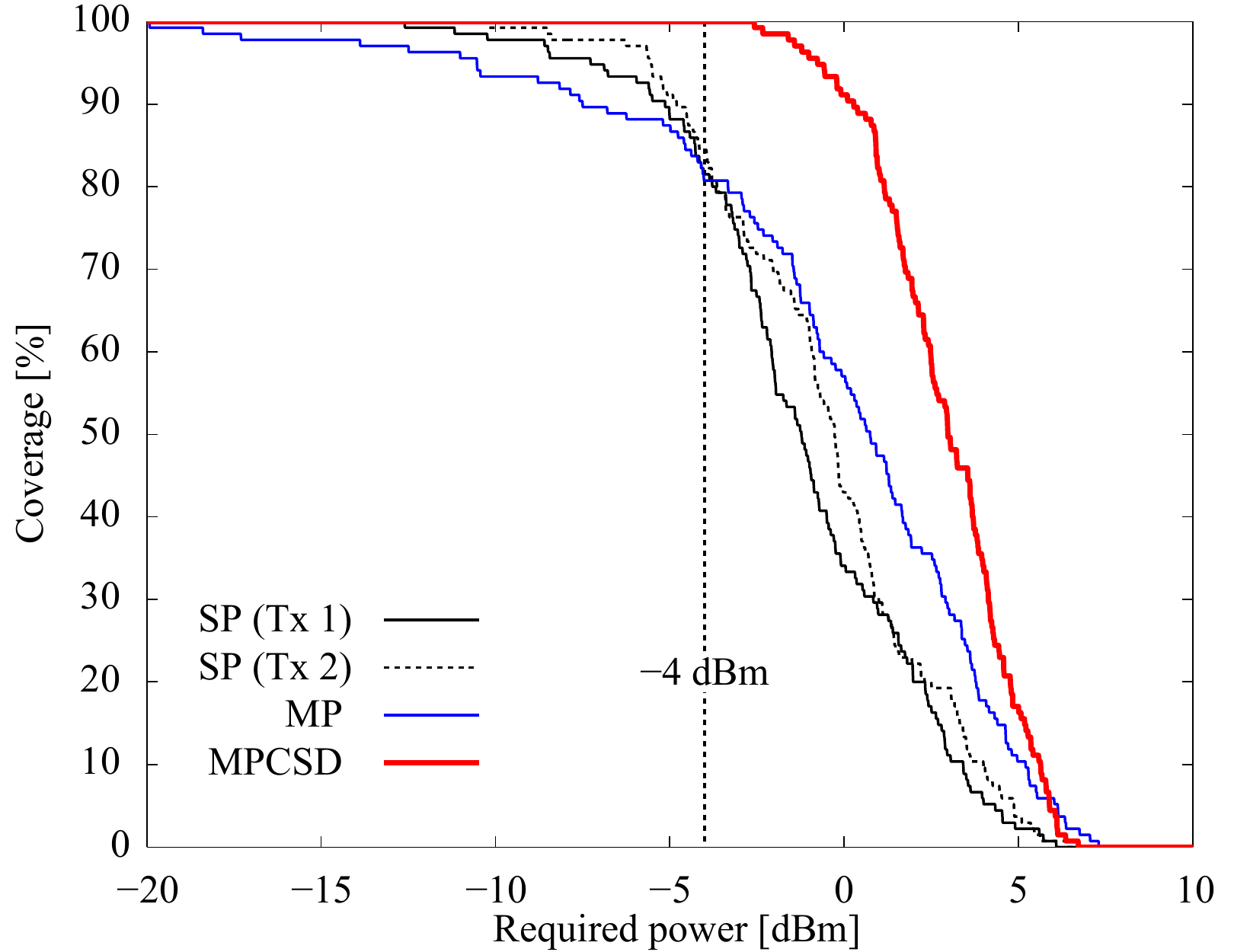}
    \caption{Coverage comparison of three energy transmission schemes against required received power.}
    \label{fig:Co}
\end{figure}

Figure~\ref{fig:Co} shows the coverage against varying $P_\mathrm{req}$. The function of the coverage can be calculated by 
\begin{eqnarray}
\hat{A} [k,P_\mathrm{req}]= \begin{cases}
1 & \mathrm{if} \ \ P_\mathrm{r}[k] \geq P_\mathrm{req} \\
0 & \mathrm{if} \ \ P_\mathrm{r}[k] < P_\mathrm{req},
\end{cases}
\end{eqnarray}
\begin{eqnarray}
\hat{C}(P_\mathrm{req}) = \frac{1}{K}\sum_{k=1}^{K}{\hat{A} [k,P_\mathrm{req}]},
\end{eqnarray}
where $P_\mathrm{r}[k]$ is the measured received power at position $k$ and $K$ is the number of measurement points, which are converted from Eqs.~(\ref{equ:eff}) and (\ref{equ:Co}).  As shown in Fig.~\ref{fig:Co}, at the required power  $P_\mathrm{req} \approx -4$~dBm, the values of coverage are almost similar to the results of the activation. From this figure, MP can only realize 100\% coverage with sensor node which consumes less than $-20$~dBm. 

\begin{table}[!t]
\caption{Measured and estimated coverage.}
\label{tab:Co}
\centering
\begin{tabular}{c||c|c|c|c}
& SP : Tx1 & SP : Tx2 & MP  & MPCSD \\
\hline
Measured& 84.4\% & 85.2\% & 83.7\% & 100\% \\
\hline
Estimated& 58.3\% & 58.3\% & 86.4\% & 100\%
\end{tabular}
\end{table}

\newpage

\section{Conclusion}
\label{sec:concl}
In this paper, we verified the effectiveness of our multi-point wireless energy transmission scheme with carrier shift diversity by the experiments using our developed battery-less sensor which requires only 400~$\mu$W with the duty cycle of 1~s. Experimental results showed that the coverage of single-point and simple multi-point energy transmission are limited to 84.4\% and 83.7\% respectively, while the proposed scheme achieves 100\% coverage. Experiments on a real battery-less sensor node verified that the improvement of the sensor activation coverage can be realized by the multi-point wireless energy transmission with carrier shift diversity.


\clearpage
\begin{thebibliography}{99}
\bibitem{solar}
D. Brunelli, C. Moser, L. Thiele, and L. Benini, ``Design of a Solar-Harvesting Circuit for Batteryless Embedded Systems,'' \emph{IEEE Trans. Circuits and Systems I}, Vol.56, No.11, pp.2519-2528, Feb. 2009.

\bibitem{RF}
U. Olgun, C. C. Chen, and J. L. Volakis, ``Design of an efficient ambient WiFi energy harvesting system,'' \emph{IET Microwaves, Antennas \& Propagation}, Vol.6, No. 11, pp.1200-1206, Aug. 2012.

\bibitem{vib}
G. K. Ottman, H. F. Hofmann, A. C. Bhatt, and G. A. Lesieutre, ``Adaptive Piezoelectric Energy Harvesting Circuit for Wireless Remote Power Supply,'' \emph{IEEE Trans. Power Electronics}, Vol.17, No.5, pp.669-676, Sep. 2002.

\bibitem{Paper0}
R. P. Wicaksono,  G. K. Tran,  K. Sakaguchi,  and K. Araki ``Wireless Grid: Enabling Ubiquitous Sensor Networks with Wireless Energy Supply'' \emph{Proc.  IEEE VTC-Spring}, pp. 1-5 May. 2011.

\bibitem{book1}
N. Shinohara, ``Power Without Wires,'' \emph{IEEE Microwave Magazine, } vol.12,  pp.S64-S73,  Dec. 2011. 

\bibitem{Paper1}
D. Maehara, G. K. Tran, K. Sakaguchi, K. Araki, M. Furukawa, ``Experiment Validating the Effectiveness of Multi-point Wireless Energy Transmission with Carrier Shift Diversity,'' \emph{IEICE Trans. Commun.}, Vol.E97-B, No.09, pp1928-1937, Sep. 2014.

\bibitem{ISSSE}
D. Maehara, G. K. Tran, K. Sakaguchi, K. Araki, T. Miyamoto, and M. Furukawa, ``Experimental Study on Multi-point Wireless Energy Transmission at 950MHz band,'' \emph{Proc. IEEE ISSSE2012},  Oct. 2012.

\bibitem{PIMRC}
D. Maehara, R. Akai, G. K. Tran, K. Sakaguchi, S. Sampei, K. Araki, H. Iwai, ``Experiment on Battery-less Sensor Activation via Multi-point Wireless Energy Transmission,'' \emph{Proc. IEEE PIMRC2013}, Sep. 2013.


\bibitem{Reg1}
``920MHz-BAND TELEMETER, TELECONTROL AND DATA TRANSMISSION RADIO EQUIPMENT,'' Assosiation of Radio Industries and Businesses ARIB STD-T108, Feb. 2012.

\bibitem{Chris}
C. R. Valenta and G. D. Durgin, ``Harvesting Wireless Power,'' \emph{IEEE Microwave Magazine, } vol.15, pp.108-120, Jun. 2014. 

\bibitem{Powercast}
\emph{Datasheet} P1110EVB, Powercast, [Online]. available: \\ http://www.powercastco.com/PDF/P1110-EVB.pdf

\bibitem{AD}
\emph{Datasheet} ADF7023, Analog Devices, [Online]. available: \\ http://www.analog.com/static/imported-files/data\_sheets\\/ADF7023.pdf


\bibitem{Smart}
D. Maehara, G. Matsushita, Y. Kuki, K. Sakaguchi, S. Sampei, K. Araki, ``[Requested Talk] Development of Battery-less Sensor Networks for LED Light Control System,'' \emph{Proc. of IEICE SmartCom2014}, Oct. 2014.

\bibitem{APMC}
G. Matsushita, D. Maehara, Y. Kuki, K. Sakaguchi, S. Sampei, K. Araki, ``Wireless Grid to realize Battery-less Sensor Networks in Indoor Environments,'' \emph{ Proc. of IEICE APMC2014}, Nov. 2014.

\bibitem{802}
 \emph{IEEE Standard for Local and metropolitan area networks - Part 15.4: Low-Rate Wireless Personal Area Networks (LR-WPANs)},
   IEEE Std. 802.15.4-2011,
   2011.
   
\bibitem{Tessera}
\emph{Datasheet} TK-RL7023+SB-L, Tessera Technology, [Online]. available: http://www.tessera.co.jp/tk-rl7023+sb.html




\bibitem{Zoya1}
E. Falkenstein, M. Roberg, and Z. Popovic, ``Low-power Wireless Power Delivery,'' \emph{IEEE Trans. on Microwave Theory and Techniques}, vol.60, no.7, pp.2277-2286, Jul. 2012.

\bibitem{Zoya2}
T. Paing, J. Shin, R. Zane, and Z. Popovic, ``Resistor Emulation Approach to Low-Power RF Energy Harvesting,'' \emph{IEEE trans. on Power Electrons}, vol.23, no.3, pp.1494-1501, Jul. 2008.


\end{thebibliography}
\end{document}